\documentclass[aps,prb,a4paper,preprint,showpacs]{revtex4}
\usepackage{amsmath}
\usepackage{amsfonts}
\usepackage{amssymb}
\usepackage{slashbox}
\usepackage{CJK}
\usepackage{graphicx}
\usepackage{amsbsy}

\begin{document}
\title{Development in the Scattering Matrix Theory:
From Spin-Orbit-Coupling Affected Shot Noise to Quantum Pumping}
\author{Rui Zhu\renewcommand{\thefootnote}{*}\footnote{Corresponding author. Electronic address:
rzhu@scut.edu.cn} }
\address{Department of Physics, South China University of
Technology, Guangzhou 510641, People's Republic of China }

\begin{abstract}
The review chapter starts by a pedagogical introduction to the
general concept of the scattering theory: from the fundamental
wave-function picture to the second-quantization language, with the
aim to clear possible ambiguity in conventional textbooks. Recent
progress in applying the method to current fluctuations and
oscillating-parameter driven quantum pumping processes is presented
with inclusion of contributions by B\"{u}ttiker, Brouwer, Moskalets,
Zhu, etc. In particular, the spin-orbit-coupling affected shot noise
can be dealt with by taking into account the spin-dependent
scattering processes. A large shot noise suppression with the Fano
factor below 0.5 observed experimentally can be illustrated by
effective repulsion between electrons with antiparallel spin induced
by the Dresselhaus spin-orbit coupling effect. A Floquet scattering
theory for quantum-mechanical pumping in mesoscopic conductors is
developed by Moskalets et al., which gives a general picture of
quantum pumping phenomenon, from adiabatic to non-adiabatic and from
weak pumping to strong pumping.

\end{abstract}


\maketitle

\tableofcontents

%\narrowtext

\newpage

\section{Pedagogical introduction to the
general concept of the scattering theory}

\subsection{Landauer-B\"{u}ttiker conductance}

The discussion is based on the scattering approach to electrical
conductance. This approach, as we will show, is conceptually simple
and transparent. Nevertheless, the generality of the scattering
approach and its conceptual clarity, make it the desired starting
point of a discussion of noise in electrical conductors. By
expanding the time-dependent scattering matrix into Fourier series,
a description of the quantum pumping phenomenon can be given.

We start with the wave function picture and consider an electron
tunneling through a one-dimensional single barrier\cite{Ref29},
which can be realized in a semiconductor heterostructure with a
layer of ${\rm{Al}}_x {\rm{Ga}}_{1 - x} {\rm{As}}$ of width $2L$
imbedded in $\rm{GaAs}$ as shown in Fig. 1. In the effective mass
approximation, the electron motion in each layer of the structure is
described by the stationary solution of the envelop equation in the
$x$-direction.
\begin{equation}
\left[ {\frac{{ - \hbar ^2 }}{2}\frac{\partial }{{\partial
x}}\frac{1}{{m^* \left( x \right)}}\frac{\partial }{{\partial x}} +
V_{eff} \left( x \right)} \right]\psi \left( x \right) = E\psi
\left( x \right).
\end{equation}
Here, $m^{*}$ and $V_{eff}$ are the effective mass and potential in
different regions with $E$ the energy of the transporting electron.
The electron's wave functions are expressible as
\begin{equation}
\Psi \left( {x,t} \right) = \psi \left( x \right)e^{ - iEt/\hbar }
= \left\{ \begin{array}{l}
 \begin{array}{*{20}c}
   {\left( {Ae^{ikx}  + Be^{ - ikx} } \right)e^{ - iEt/\hbar } ,} & {x \le -L,}  \\
\end{array} \\
 \begin{array}{*{20}c}
   {\left( {Ce^{\kappa x}  + De^{ - \kappa x} } \right)e^{ - iEt/\hbar } ,} & {-L \le x \le L}  \\
\end{array}, \\
 \begin{array}{*{20}c}
   {\left( {Ee^{ikx}  + Fe^{ - ikx} } \right)e^{ - iEt/\hbar } ,} & {x \ge L,}  \\
\end{array} \\
 \end{array} \right.
\end{equation}
with $k = {{\sqrt {2m^* E} } \mathord{\left/
 {\vphantom {{\sqrt {2m^* E} } \hbar }} \right.
 \kern-\nulldelimiterspace} \hbar }$ and $\kappa  = {{\sqrt {2m^* \left( {V_0  - E} \right)} } \mathord{\left/
 {\vphantom {{\sqrt {2m^* \left( {V_0  - E} \right)} } \hbar }} \right.
 \kern-\nulldelimiterspace} \hbar }$. The coefficients $A$ and $B$
 are associated respectively with incoming and outgoing waves on the
 left side relative to the barrier. Likewise, the coefficients $E$
 and $F$ are respectively outgoing and incoming waves on the right.
 The scattering matrix connects the incoming
 and outgoing fluxes as
\begin{equation}
 \left[ {\begin{array}{*{20}c}
   B  \\
   E  \\
\end{array}} \right] = \left[ {\begin{array}{*{20}c}
   {S_{11} } & {S_{12} }  \\
   {S_{21} } & {S_{22} }  \\
\end{array}} \right]\left[ {\begin{array}{*{20}c}
   A  \\
   F  \\
\end{array}} \right].
\end{equation}
Ideal (i.e., without scattering) conducting leads connect the
scattering region to reservoirs on the left and right characterized
by quasi-Fermi energies $\mu _1$ and $\mu _2$, respectively,
corresponding to the electron densities there. These reservoirs or
contacts randomize the phase of the injected and absorbed electrons
through inelastic processes such that there is no phase relation
between particles. For such an ideal 1D system, the current injected
from the left and right may be written as an integral over the
flux\cite{Ref29, Ref33}
\begin{equation}
I_{L} = \frac{{2e}}{{2\pi }}\left[ {\int_0^\infty  {dkv\left( k
\right)f_1 \left( k \right)T\left( E \right)}  - \int_0^\infty
{dk'v\left( {k'} \right)f_2 \left( {k'} \right)T\left( {E'} \right)}
} \right],
\end{equation}
where $v(k)$ is the velocity, $T(E)$ is the transmission
coefficient, which can be obtained from the wave function Eq. (2),
and $f_1$ and $f_2$ are the reservoir distribution functions
characterized by $\mu _1$ and $\mu _2$, respectively. The
integrations are only over positive $k$ and $k'$ relative to the
direction of the injected charge as positive $k$ is in
$+x$-direction and positive $k'$ is in $-x$-direction. If we now
assume low temperatures, electrons are injected up to an energy $\mu
_1$ from the left lead and injected up to $\mu _2$ from the right
one. Converting to integrals over energy, the current becomes
\begin{equation}
\begin{array}{l}
 I_{L} = \frac{{2e}}{{2\pi }}\left[ {\int_0^{\mu _1 } {dE\left( {\frac{{dk}}{{dE}}} \right)v\left( k \right)T\left( E \right)}  - \int_0^{\mu _2 } {dE\left( {\frac{{dk'}}{{dE}}} \right)v\left( {k'} \right)T\left( E \right)} } \right] \\
  = \frac{{2e}}{{2\pi \hbar }}\int_{\mu _2 }^{\mu _1 } {dET\left( E \right)} . \\
 \end{array}
\end{equation}
It can be seen that the first term of Eq. (5) is the flux generated
by electrons injected from the left reservoir and the second term is
that from the right reservoir. The integration is done in two
independent ensembles.

The Landauer-B\"{u}ttiker conductance introduced above can be
reproduced in the second-quantization language. Without loss of
generality, we assume the input amplitudes from the two reservoirs
in Eq. (2) to be unity. For simplicity, as a single channel is
considered, the scattering matrix can be described in the relation
\begin{equation}
\hat b_\alpha   = S_{\alpha \beta } \hat a_\beta
\end{equation}
or elaborately
\begin{equation}
\left[ {\begin{array}{*{20}c}
   {\hat b_L }  \\
   {\hat b_R }  \\
\end{array}} \right] = \left[ {\begin{array}{*{20}c}
   {S_{11} } & {S_{12} }  \\
   {S_{21} } & {S_{22} }  \\
\end{array}} \right]\left[ {\begin{array}{*{20}c}
   {\hat a_L }  \\
   {\hat a_R }  \\
\end{array}} \right] = \left[ {\begin{array}{*{20}c}
   r & {t'}  \\
   t & {r'}  \\
\end{array}} \right]\left[ {\begin{array}{*{20}c}
   {\hat a_L }  \\
   {\hat a_R }  \\
\end{array}} \right],
\end{equation}
where operators $a_{L/R}$ annihilate electrons incident upon the
sample from the left/right reservoir and operators $b_{L/R}$
describe electrons in the outgoing states. $t$ and $t'$ are the
transmission amplitudes of electrons incident rightward and
rightward, respectively, and $r$ and $r'$ are the corresponding
reflection amplitudes, as defined conventionally. Hence, the
Landauer-B\"{u}ttiker formula of the current can be expressed as
\begin{equation}
I_L  = \frac{e}{{2\pi \hbar }}\int {dE\left\langle {\left[ {a_L^\dag
\left( E \right)a_L \left( E \right) - b_L^\dag  \left( E \right)b_L
\left( E \right)} \right]} \right\rangle } ,
\end{equation}
where $\left\langle  \cdots  \right\rangle $ calculates the quantum
statistical average of the product of an electron creation operator
and annihilation operator of a Fermi gas. As a conventional electron
reservoir is considered, we have
\begin{equation}
\begin{array}{l}
 \left\langle {a_L^\dag  \left( E \right)a_L \left( {E'} \right)} \right\rangle  = f_L \left( E \right)\delta \left( {E - E'} \right), \\
 \left\langle {a_R^\dag  \left( E \right)a_R \left( {E'} \right)} \right\rangle  = f_R \left( E \right)\delta \left( {E - E'} \right), \\
 \left\langle {a_L^\dag  \left( E \right)a_R \left( {E'} \right)} \right\rangle  = \left\langle {a_R^\dag  \left( E \right)a_L \left( {E'} \right)} \right\rangle  = 0. \\
 \end{array}
\end{equation}
It should be noted from the third formula of Eq. (9) that electrons
incident from the left and right reservoirs are completely
incoherent and the phases of the two reservoirs are randomized and
completely unrelated. Substituting Eq. (9) to Eq. (8), the formula
of the current expressed in Eq. (5) can be reproduced.

It is interesting to see that if we could build up a system with the
conductor connected to two correlated reservoirs, in which the
quantum statistical average of the cross product reads
\begin{equation}
\begin{array}{l}
 \left\langle {a_L^\dag  \left( E \right)a_R \left( {E'} \right)} \right\rangle  = \left\langle {a_R^\dag  \left( E \right)a_L \left( {E'} \right)} \right\rangle  = f_3 \left( E \right)\delta \left( {E - E'} \right), \\
 f_3 \left( E \right) = \left\{ \begin{array}{l}
 1,\begin{array}{*{20}c}
   {} & {\mu  < \min \left( {\mu _1 ,\mu _2 } \right),}  \\
\end{array} \\
 0,\begin{array}{*{20}c}
   {} & {{\rm{others}},}  \\
\end{array} \\
 \end{array} \right. \\
 \end{array}
\end{equation}
the system can generate a current demonstrating the interference
between the electron states incident from the left reservoir and the
right one.
\begin{equation}
I_L  =  - \frac{e}{{2\pi \hbar }}\int_0^{\mu _1 } {dE} \left[
{\int_0^{\mu _1 } {\left| r \right|^2 dE}  + \int_0^{\mu _2 }
{\left| t \right|^2 dE}  + 2\int_0^{\mu \left( {\min } \right)}
{{\mathop{\rm Re}\nolimits} \left( {r^* t'} \right)dE} } \right].
\end{equation}
In the next subsection, we would use a toy system based on
correlated reservoirs to further illustrate the scattering scheme.
Significant difference between uncorrelated and correlated
reservoirs is illuminated. It is also elaborated that in scattering
problems the transmission and reflection process demonstrates the
quantum state spanned throughout the space.

\subsection{Further illustration of the scattering scheme
in a toy system with correlated reservoirs}

We consider incidence only from the left in Eq. (2) and the input
flux is assumed to be unity. The wave function in different
scattering regions can be expressed as
\begin{equation}
\psi \left( x \right) = \left\{ \begin{array}{l}
 \begin{array}{*{20}c}
   {e^{ikx}  + re^{ - ikx} ,} & {x \le  - L,}  \\
\end{array} \\
 \begin{array}{*{20}c}
   {Ce^{\kappa x}  + De^{ - \kappa x} ,} & { - L \le x \le L}  \\
\end{array}, \\
 \begin{array}{*{20}c}
   {te^{ikx} ,} & {x \ge L.}  \\
\end{array} \\
 \end{array} \right.
\end{equation}
An incident particle is transmitted with probability $\left| t
\right|^2 $ and reflected with probability $\left| r \right|^2 $.
 It is determined by the wave function that
 the particle momentum has value of $\hbar k$
with probability ${{\left( {1 + \left| t \right|^2 } \right)}
\mathord{\left/
 {\vphantom {{\left( {1 + \left| t \right|^2 } \right)} 2}} \right.
 \kern-\nulldelimiterspace} 2}$
 and $-\hbar k$ with probability ${{\left| r \right|^2 } \mathord{\left/
 {\vphantom {{\left| r \right|^2 } 2}} \right.
 \kern-\nulldelimiterspace} 2}$. The mean value of the momentum
 $\left| t \right|^2 \hbar k$
characterizes the density of the probability current $\bf{J}$, which
can also be obtained from the continuity equation $ {{\partial \rho
} \mathord{\left/
 {\vphantom {{\partial \rho } {\partial {\rm{t}}}}} \right.
 \kern-\nulldelimiterspace} {\partial {\rm{t}}}} + \nabla  \cdot {\bf{J}} = 0
$ with $\rho$ the probability density.

An interesting interference pattern can be observed if we consider
the incidence from the left and the right is correlated. We consider
a toy system of a one-dimensional electron gas subject to two
oscillating gate volatages (see Fig. 2). The inspiration comes from
the quantum pumping phenomenon. The single-particle Hamiltonian
reads
\begin{equation}
H =  - \frac{\hbar }{{2m^{*}}}\frac{{\partial ^2 }}{{\partial x^2 }}
+ U\left( {x,t} \right),
\end{equation}
with $U\left( {x,t} \right) = \Theta \left( {x + 2L} \right)\Theta
\left( { - L - x} \right)U_1 \left( t \right) + \Theta \left( {x -
L} \right)\Theta \left( {2L - x} \right)U_2 \left( t \right).$ The
two barriers $U_1$ and $U_2$ are adiabatically modulated at the
frequency $\omega$ with a phase difference $\phi$.
\begin{equation}
\begin{array}{l}
 U_1 \left( t \right) = U_{10}  + U_{1\omega } \sin \omega t, \\
 U_2 \left( t \right) = U_{20}  + U_{2\omega } \sin \left( {\omega t - \phi } \right). \\
 \end{array}
\end{equation}
We assume $\omega$ is extremely small so that a static treatment is
valid. Therefore, it is tolerable to consider only the zero order of
the Fourier component of the time-dependent scattering matrix
without taking into account the photon-absorption/emmission
processes, which is done in adiabatic quantum pumping theory.

As shown in Fig. 2, the electrons are incident from the left and
right reservoirs with identical amplitudes at zero bias, which is
set to be unity without impairing generality. As assumed, incidence
from the two correlated reservoirs characterize a coherent
single-particle state. The single-particle wave function at a
certain time has the following form.
\begin{equation}
\Psi \left( {x,t} \right) = \left\{ \begin{array}{l}
 \begin{array}{*{20}c}
   {\left[ {e^{ikx}  + \left( {r + t'e^{i\theta } } \right)e^{ - ikx} } \right]e^{{{ - iEt} \mathord{\left/
 {\vphantom {{ - iEt} \hbar }} \right.
 \kern-\nulldelimiterspace} \hbar }} ,} & {x \le  - 2L,}  \\
\end{array} \\
 \begin{array}{*{20}c}
   {\left[ {A_2 e^{\kappa _2 x}  + B_2 e^{ - \kappa _2 x} } \right]e^{{{ - iEt} \mathord{\left/
 {\vphantom {{ - iEt} \hbar }} \right.
 \kern-\nulldelimiterspace} \hbar }} ,} & { - 2L \le x \le  - L}  \\
\end{array}, \\
 \begin{array}{*{20}c}
   {\left[ {A_3 e^{ikx}  + B_3 e^{ - ikx} } \right]e^{{{ - iEt} \mathord{\left/
 {\vphantom {{ - iEt} \hbar }} \right.
 \kern-\nulldelimiterspace} \hbar }} ,} & { - L \le x \le L}  \\
\end{array}, \\
 \begin{array}{*{20}c}
   {\left[ {A_4 e^{\kappa _4 x}  + B_4 e^{ - \kappa _4 x} } \right]e^{{{ - iEt} \mathord{\left/
 {\vphantom {{ - iEt} \hbar }} \right.
 \kern-\nulldelimiterspace} \hbar }} ,} & {L \le x \le 2L}  \\
\end{array}, \\
 \begin{array}{*{20}c}
   {\left[ {\left( {t + r'e^{i\theta } } \right)e^{ikx}  + e^{ - ikx} } \right]e^{{{ - iEt} \mathord{\left/
 {\vphantom {{ - iEt} \hbar }} \right.
 \kern-\nulldelimiterspace} \hbar }} ,} & {x \ge 2L.}  \\
\end{array} \\
 \end{array} \right.
\end{equation}
$k = {{\sqrt {2m^* E} } \mathord{\left/
 {\vphantom {{\sqrt {2m^* E} } \hbar }} \right.
 \kern-\nulldelimiterspace} \hbar }$ and $\kappa_{2/4}  = {{\sqrt {2m^* \left( {U_{1/2}
   - E} \right)} } \mathord{\left/
 {\vphantom {{\sqrt {2m^* \left( {U_{1/2}  - E} \right)} } \hbar }} \right.
 \kern-\nulldelimiterspace} \hbar }$.
$t$ and $r$ quantify the transmission and reflection amplitudes of
the electrons incident from the left reservoir while $t'$ and $r'$
quantify those incident from the right with $t' = t$ and $r' = {{ -
tr^* } \mathord{\left/
 {\vphantom {{ - tr^* } {t^* }}} \right.
 \kern-\nulldelimiterspace} {t^* }}$
. We introduce a geometric phase $\theta$ to describe the
unavoidable phase difference between the electrons injected from the
two reservoirs. In the toy approach, correlated reservoirs are
assumed, which justifies a particular value of $\theta$. For
simplicity and without violation of the physics we take $\theta =0$.
It is noted that in conventional real reservoirs mixed-ensemble
integral should be applied, i.e., the probability flow of the
electron incident from the left reservoir absolutely cancels out
that of the incidence backward at zero bias as a result of the
randomized phase distribution, which absolutely differs from our toy
consideration.

From the continuity equation
\begin{equation}
\frac{{\partial \rho }}{{\partial t}} + \nabla  \cdot {\bf{J}} = 0,
\end{equation}
we could derive the probability current flow as functions of the
transmission and reflection amplitudes as
\begin{equation}
j_x  =  - \frac{{i\hbar }}{{2m}}\left[ {\psi ^\dag  \frac{\partial
}{{\partial x}}\psi  - \left( {\frac{\partial }{{\partial x}}\psi
^\dag  } \right)\psi } \right]=
 - \frac{{4\hbar k}}{m}{\mathop{\rm Re}\nolimits} \left( {r^* t'}
 \right).
\end{equation}
We can also see from the wave function Eq. (15) that the particle
momentum has value of $\hbar k$ with probability ${{\left[ {1 -
{\mathop{\rm Re}\nolimits} \left( {tr^* } \right)} \right]}
\mathord{\left/
 {\vphantom {{\left[ {1 - {\mathop{\rm Re}\nolimits} \left( {tr^* } \right)} \right]} 2}} \right.
 \kern-\nulldelimiterspace} 2}$
and $-\hbar k$ with probability ${{\left[ {1 + {\mathop{\rm
Re}\nolimits} \left( {tr^* } \right)} \right]} \mathord{\left/
 {\vphantom {{\left[ {1 + {\mathop{\rm Re}\nolimits} \left( {tr^* } \right)} \right]} 2}} \right.
 \kern-\nulldelimiterspace} 2}$. The mean value of the momentum
 $ - {\mathop{\rm Re}\nolimits} \left( {tr^* } \right)\hbar k$
characterizes the density of the probability current of Eq. (17).
The net current density can be described by the period-average of
the probability current density multiplied by the carrier charge and
density. The accumulated contribution by electrons within the $ \pm
\hbar \omega $ sidebands is taken into account by an integral.
Without dynamic modulation, no current occurs as the $ \pm \hbar
\omega $ energy channel is closed even when the probability current
density is nonzero for asymmetric barrier configuration. The current
density as a function of the Fermi level thus becomes
\begin{equation}
I_{L}\left( E_{F} \right) =  - \frac{{4\hbar }}{{m^* }}\int_{E_F  -
\hbar \omega }^{E_F  + \hbar \omega } {\frac{{e k \rho _e N\left( E
\right)f\left( E \right)}}{{\left( {{{2\pi } \mathord{\left/
 {\vphantom {{2\pi } \omega }} \right.
 \kern-\nulldelimiterspace} \omega }} \right)}}\int_0^{\frac{{2\pi }}{\omega }} {{\mathop{\rm Re}\nolimits} \left[ {r^* \left( t \right)t'\left( t \right)} \right]dt} dE}
,
\end{equation}
where the density of states of a one-dimensional electron gas is
\begin{equation}
N\left( E \right) = \frac{V}{{\pi \hbar }}\sqrt {\frac{{m_e
}}{{2E}}}.
\end{equation}
Here, $k$ is the wave vector of the electron. $e$ is the electron
charge. $\rho _e $ is the carrier density of the two-dimensional
electron gas in which the quantum wire is confined. $f(E)$ is the
Fermi-Dirac distribution function of the leads. $V$ quantifies the
volume of the one-dimensional wire and $m_e$ is the mass of a free
electron.

Here, in the toy configuration, quantum interference is remarkably
demonstrated. The single-electron state Eq. (15) interferes with
itself and carries the probability flow and hence the net current
through oscillating cycles.

We numerically calculated the current in a one-dimensional electron
gas (2DEG) based on the GaAs/AlGaAs heterostructures with the
average carrier density $\rho _e \sim 10^{11} $ cm$^{-2}$ and the
effective mass of the electron $m^* \sim 0.067m_e $\cite{Ref73}. The
width of the two gate potential barriers $L=20$ {\AA} equally
separated by a $2L=40$ {\AA} width well. The amplitudes of the
modulations $U_{1\omega }  = U_{2\omega }  = 1$ meV. All of the
above setups are not essential as we are dealing with an assumed toy
structure.

In Fig. 3, it is shown that the time-integrated current demonstrates
a sinusoidal pattern as a function of the phase difference between
two oscillating parameters, which is a result of quantum
interference. The probability density flow formulated in Eq. (17) is
a result of phase difference between transmission forward and
backward. Quantum phase interference gives rise to nonzero
probability density flow for asymmetric barrier configurations. The
absolute strength of the probability flow increases as the height
difference between the two potential barriers increases determined
by the phase difference $\phi$. Fig. 4 presented the time variation
of the net current within an oscillating cycle. Considering the
time-averaged effect, the integrated asymmetry is different. When
the phase difference between the two modulations approaches $\pi$,
time-reversal symmetry destroys the time-integrated current to zero
although the probability density flow maximizes at a certain time.
i.e. the probability flow in half a pumping cycle completely offsets
that of the other. Therefore a sinusoidal dependence on $\phi$
occurs in the time-integrated current.

In real reservoirs, the single-particle state between reservoirs has
a definite momentum direction determined by its source. When
coherent reservoirs can be realized in any form, however, quantum
states within the mesoscopic conductor can be expressed as Eq. (15)
and double-slit interference pattern is observable in an electron
device.

\subsection{Time-dependent scattering-matrix theory}

The scattering-matrix equation $\hat b_\alpha   = S_{\alpha \beta }
\hat a_\beta  $ with $\alpha$ and $\beta$ indexes of lead, channel,
and spin introduced in Sec. I.A. for the Landauer-B\"{u}ttiker
conductance characterizes the transport properties through a
conductor at a certain bias.

In a more general situation with dynamic processes, e.g. in quantum
pumping, a time-dependent scattering matrix can be introduced as
follows.
\begin{equation}
\hat b_\alpha  \left( t \right) = \int_{ - \infty }^\infty
{S_{\alpha \beta } \left( {t,t'} \right)\hat a_\beta  \left( {t'}
\right)dt'} ,\begin{array}{*{20}c}
   {} & {t \ge t',}  \\
\end{array}
\end{equation}
with $\alpha$ and $\beta$ general indexes denoting the lead,
channel, and spin. An incident state $\hat a_\beta$ at time $t'$ is
scattered into the outgoing state $\hat b_\alpha$ at time $t$ with
the amplitude $S_{\alpha \beta } \left( {t,t'} \right)$.

The time-dependent scattering-matrix picture described by Eq. (20)
is exactly equivalent to the time-dependent Schr\"{o}dinger equation
with the elements of the scatting matrix amplitudes of the wave
function. In usual cases, the time-dependent Schr\"{o}dinger
equation cannot be solved exactly, similarly to the time-dependent
scattering matrix. In the static or adiabatic cases, it is
advantageous to use an analog of the Wigner transform for the matrix
$S_{\alpha \beta } \left( {t,t'} \right)$\cite{trs},
\begin{equation}
S_{\alpha \beta } \left( {E,t} \right) = \int_{ - \infty }^\infty
{e^{iE\left( {t - t'} \right)} S_{\alpha \beta } \left( {t,t'}
\right)dt'}.
\end{equation}
An on-time scattering process $S_{\alpha \beta } \left( {t,t'}
\right)$ with $t=t'$ is sufficient to describe the bias-driven
conductance. Namely, from Eq. (21), we can use $S_{\alpha \beta }
\left( {E} \right)$ with $E$ labeling the energy channel to fully
capture the transport physics.

When the scattering time $t-t'$ is small (i.e., the dynamic
characteristic frequency is much smaller than the inverse Wigner
time delay), the dynamics can be approximated into the
instant-scattering picture. Physically this means that the
scattering matrix changes only a little while an electron is
scattered by the mesoscopic sample under dynamic modulation, in
which we use the term ``adiabatic". In adiabatic dynamics, we can
use low-order Fourier components of $S_{\alpha \beta } \left( {E,t}
\right)$ to characterize transport physics. Therefore, small $t-t'$
is transformed into variation of the particle energy by side-band
broadening around the Fermi level.

In Sec. III, we would illustrate the time-dependent scattering
approach in adiabatic quantum pumping beyond the linear-response
approximation.

\section{Spin-orbit coupling affected shot noise}

\subsection{Background}

Current fluctuations are present in almost all kinds of conductors
and have been developed into a very active and fascinating subfield
of mesoscopic physics (for review see Ref. \onlinecite{Ref20}). At
low temperatures, thermal fluctuations are extremely small, the
current fluctuation properties are governed by the so-called shot
noise, which is a consequence of the quantization of charge. Shot
noise is useful to obtain information on a system which is not
available through conductance measurements. In particular, shot
noise experiments can determine the quantum correlation of
electrons, the charge and statistics of the quasi-particles relevant
for transport, and reveal information on the potential profile as
well as internal energy scales of mesoscopic systems\cite{Ref34,
Ref35, Ref36, Ref37, Ref38, Ref39, Ref40, Ref41, Ref42, Ref43,
Ref44, Ref45, Ref22, Ref60}. Shot noise is generally more sensitive
to the effects of electron-electron interactions than the average
conductance.

A convenient measure of shot noise is the Fano factor $F$, which is
the ratio of the actual shot noise and the Poisson noise. The
Poisson noise would be achieved in measurement if the transport is
carried by single independent electrons. Four typical values of the
Fano factor characterize the shot noise properties of different
mesoscopic conductors.

1. $F=1$ characterizes Poissonian processes. Particles are
completely independent during transport corresponding to channels
through which transmission is exponentially small. In diffusive
transport, they are the so-called approximately-closed channels.
Typical conductors featuring $F=1$ include tunneling junction,
Schottky-barrier diode, and asymmetric double-barrier diode.

2. $F=0$ characterizes ballistic transport. In ballistic transport,
transmission approaches the maximum of unity. Free particles wave
function extends throughout the space, i.e. particle beams exhibit
full coherence. In diffusive transport, they are the so-called open
channels. Typical conductors featuring $F=0$ include pure metal and
free two-dimensional electron gas.

3. $F=1/3$ characterizes diffusive transport. Open channels and
closed channels are distributed randomly. As a result of ensemble
average of channels, the strength shot noise falls between $F=1$ and
$F=0$. Typical conductors featuring $F=1/3$ include diffusive
metals, graphene, and two-dimensional electron gas modulated by
magnetic barriers.

4. $F=1/2$ characterizes ballistic transport constrained by Pauli
principle. The Pauli principle forbids two electrons to be in the
same channel simultaneously. As a result, the shot noise is
suppressed. All kinds of conductors are subject to the Pauli
principle. In the symmetric-double-barrier diode, where Pauli
exclusion is the only correlation between particles, the shot noise
features $F=1/2$.

With increased attention\cite{Ref46, Ref47, Ref48, Ref49, Ref50,
Ref51} to semiconductor spintronics, materials such as $ {\rm{GaSb}}
$, $ {\rm{InAs}} $, and $ {\rm{InSb}} $ with considerably strong
spin-orbit coupling (SOC) constant\cite{Ref51, Ref52} are becoming
widely used in mesoscopic conductors, observations\cite{Ref34,
Ref37, Ref38} beyond the formalism of earlier theory on current shot
noise are reported. The scattering approach is developed to derive a
general formula for the shot noise in the presence of the SOC effect
and apply it to the double-barrier resonant diode (DBRD) systems. It
is demonstrated that the microscopic origin of the super-suppression
of the shot noise observed in experiment\cite{Ref34, Ref37, Ref38}
is the bunching interaction between electrons with opposite spins
resulting from the Dresselhaus $k^{3}$ terms\cite{Ref53, Ref54} in
the effective Hamiltonian of the bulk semiconductor of the barriers.

\subsection{Theoretical approach}

In this part, the effect of the Dresselhaus SOC to the shot noise
properties in the DBRD structure connected to ferromagnetic or
normal metal leads is considered. The theory can be generalized to
other coherent mesoscopic conductors subject to Dresselhaus and/or
Rashba SOC.

The Dresselhaus SOC is caused by the bulk inversion asymmetry and
exists broadly in III-V compound semiconductors with zinc-blende
crystal structures\cite{Ref53}. Consider the transmission of
electrons with identical wave vector $ {\bf{k}} =
({\bf{k}}_\parallel  ,k_z ) $ through a certain potential barrier
grown along $z\parallel [001]$ direction. ${\bf{k}}_\parallel$ is
the wave vector in the plane of the barrier and $k_{z}$ is the wave
vector component normal to the barrier. The electron Hamiltonian of
the barrier in the effective-mass approximation contains the
spin-dependent $k^{3}$ Dresselhaus term
\begin{equation}
\hat H_D  = \gamma (\hat \sigma _x k_x  - \hat \sigma _y k_y
)\frac{{\partial ^2 }}{{\partial z^2 }},
\end{equation}
where $\gamma$ is the material constant denoting the strength of the
Dresselhaus SOC, $\hat \sigma _x $ and $\hat \sigma _y $ are the
Pauli matrices. The Dresselhaus term can be diagonalized by the
spinors
\begin{equation}
\chi _ \pm   = \frac{1}{{\sqrt 2 }}\left( {\begin{array}{*{20}c}
   1  \\
   { \mp e^{ - i\varphi } }  \\
\end{array}} \right),
\end{equation}
which describe the spin-up (``$+$") and spin-down (``$-$") electron
eigenstates.

Suppose the system is a layered mesoscopic conductor with its
potential profile described by $V_{0} (z)$ (see Fig. 5), the
electron motion in each layer of the structure is described by the
Hamiltonian
\begin{equation}
 \hat H = - \frac{{\hbar ^2 }}{{2m^* }}\nabla ^2  +
V(z) + \hat H_D,
\end{equation}
where $ V(z) = V_0 (z) - eF(z + b)\Theta (z + b)\Theta (a + c - z) $
with $F$ the magnitude of the electric field, $\Theta (z)$ the step
function, and $-b$ and $a+c$ the longitudinal coordinates of
surfaces in $z$ direction. Our discussion is within the framework of
single electron approximation and coherent tunneling\cite{Ref55,
Ref56}, and only zero-frequency noise at zero temperature is
considered. Under the assumption that $ k_\parallel $ is conserved
during the tunneling, the wave functions for the electrons with
definite longitudinal electron energy ($E_{z}$) can be obtained from
Schr\"{o}dinger equation based on the Hamiltonian given in Eq. (24),
which can be diagonalized by spinors $ \chi _ \pm $. So, the wave
functions become
\begin{equation}
\psi
(\mathord{\buildrel{\lower3pt\hbox{$\scriptscriptstyle\rightharpoonup$}}
\over r} ) = \left\{ {\begin{array}{*{20}c}
   {\exp (i{\bf{k}}_\parallel   \cdot {\bf{\rho }})\left( {\sum\limits_{j =  \pm } {\sqrt {\frac{{m_1 }}{{\hbar k_{1j} }}} \exp (ik_{1j} z)\chi _j }  + \sum\limits_{j =  \pm } {r_j \sqrt {\frac{{m_1 }}{{\hbar k_{1j} }}} \exp ( - ik_{1j} z)\chi _j } } \right) ,} & {z < b ,}  \\
   {\exp (i{\bf{k}}_\parallel   \cdot {\bf{\rho }})\sum\limits_{j =  \pm } {\Phi _{\xi j} (z)} ,} & { - b \le z < a + c ,}  \\
   {\exp (i{\bf{k}}_\parallel   \cdot {\bf{\rho }})\sum\limits_{j =  \pm } {t_j \sqrt {\frac{{m_5 }}{{\hbar k_{5j} }}} \exp (ik_{5j} z)\chi _j } ,} & {z \ge a + c ,}  \\
\end{array}} \right.
\end{equation}
where $ \Phi _{\xi j} $ denotes the wave function in the conductor
region, and $t_{j}$ and $r_{j}$ are the transmission and reflection
amplitudes, which can be calculated using the transfer-matrix
method\cite{Ref57}. The spin ``$+$" and spin ``$-$" components of
the electron wave functions transport separately without
correlation. As standard scattering method is applied, we introduce
creation and annihilation operators of electrons in the scattering
states. Schematics of the scattering states are shown in Fig. 6.
Operators $ \hat a_{Ln\sigma }^\dag (E) $ and $ \hat a_{Ln\sigma }
(E) $ create and annihilate electrons with total energy $E$ and spin
polarization $\sigma$ in the transverse channel $n$ in the left
lead, which are incident upon the sample. In the same way, the
creation $ \hat b_{Ln\sigma }^\dag (E) $ and annihilation $ \hat
b_{Ln\sigma } (E) $ operators describe electrons in the outgoing
states. They obey anti-commutation relations. Therefore, we can
write the scattering matrix of the sample as
\begin{equation}
\begin{array}{c}
 \left( {\begin{array}{*{20}c}
   {\hat b_{Ln \uparrow } }  \\
   {\hat b_{Ln \downarrow } }  \\
   {\hat b_{Rn \uparrow } }  \\
   {\hat b_{Rn \downarrow } }  \\
\end{array}} \right) = \frac{1}{2}\underbrace {\left( {\begin{array}{*{20}c}
   1 & 1 & 0 & 0  \\
   {e^{ - i\varphi } } & { - e^{ - i\varphi } } & 0 & 0  \\
   0 & 0 & 1 & 1  \\
   0 & 0 & {e^{ - i\varphi } } & { - e^{ - i\varphi } }  \\
\end{array}} \right)}_{M_1 } \times \underbrace {\left( {\begin{array}{*{20}c}
   {r_ +  } & 0 & {t_ +  '} & 0  \\
   0 & {r_ -  } & 0 & {t_ -  '}  \\
   {t_ +  } & 0 & {r_ +  '} & 0  \\
   0 & {t_ -  } & 0 & {r_ -  '}  \\
\end{array}} \right)}_s \\
  \times \underbrace {\left( {\begin{array}{*{20}c}
   1 & { - e^{i\varphi } } & 0 & 0  \\
   1 & {e^{i\varphi } } & 0 & 0  \\
   0 & 0 & 1 & { - e^{i\varphi } }  \\
   0 & 0 & 1 & {e^{i\varphi } }  \\
\end{array}} \right)}_{M_2 } \times \left( {\begin{array}{*{20}c}
   {\hat a_{Ln \uparrow } }  \\
   {\hat a_{Ln \downarrow } }  \\
   {\hat a_{Rn \uparrow } }  \\
   {\hat a_{Rn \downarrow } }  \\
\end{array}} \right) ,\\
 \end{array}
\end{equation}
with $ t' = t,r' =  - \frac{t}{{t^* }}r^* $, $ T_j (E) = \left| {t_j
} \right|^2 ,R_j (E) = \left| {r_j } \right|^2 $, and $j =  \pm $.
The matrices $M_{1}$ and $M_{2}$ are unitary transformations between
spin ``$\uparrow\downarrow$" states and spin ``$\pm$" states, $s$ is
the scattering matrix connecting the incoming and outgoing spin
``$\pm$" states of the $n$th channel. The current of the system can
be derived as follows
\begin{equation}
\hat I_L (t) = \frac{e}{{2\pi \hbar }}\sum\limits_{\alpha \beta }
{\sum\limits_{mn} {\sum\limits_{\sigma '\sigma ''}
{\sum\limits_\sigma  {\int {dE} \int {dE'} e^{i(E - E')t/\hbar }
\hat a_{\alpha m\sigma '}^\dag  } } } } (E)A_{\alpha ,\beta ,\sigma
',\sigma ''}^{m,n,\sigma } (L;E,E')\hat a_{\beta n\sigma ''} (E'),
\end{equation}
where
\begin{equation}
\begin{array}{l}
 A_{\alpha ,\beta ,\sigma ',\sigma ''}^{m,n,\sigma } (L;E,E') =  \\   \delta _{\alpha L} \delta _{mm} \delta _{\beta L} \delta _{\sigma '\sigma } \delta _{\sigma ''\sigma }
  - \sum\limits_{\xi '\xi ''} {\sum\limits_k {M_{2\xi ''\sigma '}^\dag  s_{L,\alpha ;m,k;\xi ''}^\dag  } (E)M_{1\sigma \xi ''}^\dag  M_{1\sigma \xi '} s_{L,\beta ;k,n;\xi '} (E')M_{2\xi '\sigma ''} } . \\
 \end{array}
\end{equation}

For a system at thermal equilibrium, the quantum statistical average
of the product of an electron creation operator and annihilation
operator of a Fermi gas with spin polarization is
\begin{equation}
\begin{array}{*{20}c}
   {\left\langle {a_{Ln \uparrow }^\dag  a_{Ln \uparrow } } \right\rangle  = f_{Lp} ,} & {\left\langle {a_{Ln \downarrow }^\dag  a_{Ln \downarrow } } \right\rangle  = f_{La} ,} & {\left\langle {a_{Ln \uparrow }^\dag  a_{Ln \downarrow } } \right\rangle  = \left\langle {a_{Ln \downarrow }^\dag  a_{Ln \uparrow } } \right\rangle  = \frac{1}{2}(f_{Lp}  - f_{La} ).}  \\
\end{array}
\end{equation}
Without loss of generality, we set the unit vector directed along
the spin orientation $ {\bf{n}}_s  = (1,0,0) $. Making use of Eqs.
(24)-(26), after some algebra, we can obtain the expression for the
zero-frequency noise power
\begin{equation}
\begin{array}{l}
 S_{\alpha \beta }  \equiv S_{\alpha \beta } (0) = \frac{{e^2 }}{{4\pi \hbar }}\sum\limits_{\delta \delta _2 } {\sum\limits_{mn} {\sum\limits_{\scriptstyle \sigma '\sigma '' \hfill \atop
  \scriptstyle \sigma _2 '\sigma _2 '' \hfill} {\sum\limits_{\scriptstyle \sigma  \hfill \atop
  \scriptstyle \sigma _2  \hfill} {\int {dE} } } } }  \\
  \times \left\{ {A_{\delta ,\delta _2 ,\sigma ',\sigma ''}^{m,n,\sigma } (\alpha ;E,E)A_{\delta _2 ,\delta ,\sigma _2 ',\sigma _2 ''}^{n,m,\sigma _2 } (\beta ;E,E) + A_{\delta ,\delta _2 ,\sigma ',\sigma ''}^{m,n,\sigma } (\beta ;E,E)A_{\delta _2 ,\delta ,\sigma _2 ',\sigma _2 ''}^{n,m,\sigma _2 } (\alpha ;E,E)} \right\} \\
  \times \left\{ {\left[ {\delta _{\sigma ''\sigma _2 '} \delta _{\sigma '\sigma _2 ''} f_\delta  (E) + \delta _{\sigma ''\sigma _2 '} \delta _{\bar \sigma '\sigma _2 ''} \frac{1}{2}\left( {f_{p\delta } (E) - f_{a\delta } (E)} \right)} \right]} \right. \\
 \left. { - \left[ {\delta _{\sigma '\sigma _2 ''} f_\delta  (E) + \delta _{\bar \sigma '\sigma _2 ''} \frac{1}{2}\left( {f_{p\delta } (E) - f_{a\delta } (E)} \right)} \right] \times \left[ {\delta _{\sigma ''\sigma _2 '} f_{\delta _2 } (E) + \delta _{\bar \sigma ''\sigma _2 '} \frac{1}{2}\left( {f_{p\delta _2 } (E) - f_{a\delta _2 } (E)} \right)} \right]} \right\}. \\
 \end{array}
\end{equation}
Eq. (30) gives a general formula to calculate the shot noise in
mesoscopic conductors subject to the SOC effect. It can be used to
predict the low-frequency noise properties of arbitrary
multi-channel, multi-probe phase-coherent conductors in the presence
of Dresselhaus SOC. It can be naturally extended to the system with
Rashba SOC and the system with both Dresselhaus SOC and Rashba SOC.
In the limit of zero SOC and when the conductor is connected to
normal metal leads, there is $ f_{\alpha a} (E) = f_{\alpha p} (E)$
and $T_{n + } (E) = T_{n - } (E) = T(E) $, thus Eq. (30) reconverts
to the formula provided by B\"{u}ttiker\cite{Ref58} concerning
scalar electron systems without the spin degree of freedom inducing
observable difference.

\subsection{Numerical results in comparison with experiment}

To further demonstrate our theory, we provide numerical results
based on real heterostructures in comparison with experiment.

The double-barrier structure considered here is constructed of
layers of $ {\rm{Ga}}_{\rm{x}} {\rm{Al}}_{{\rm{1 - x}}} {\rm{Sb}} $
with $x=0.15/0.3/0/0.3/0.15$ ($x=0.3$ for the barriers and $x=0$ for
the well), which are known to be semiconductors with relatively
strong Dresselhaus SOC\cite{Ref55, Ref56}. Our target setup is a
symmetric double-barrier structure with the thickness of the well
$a=30$ ${\rm{{\AA}}}$ and the thickness of the two barriers $b=c=50$
${\rm{{\AA}}}$. The height of the barrier $V_{b}=230$ ${\rm {meV}}$
and the depth of the well $V_{w}=200$ ${\rm {meV}}$ are given by the
heterostructure properties\cite{Ref57} (cf. Fig. 5). We assume that
in the whole region the effective mass
$m^{*}=0.053m_{e}$\cite{Ref55, Ref56}. The chemical potential of the
two electrodes is set to be ${\rm {12}}$ ${\rm {meV}}$. For
comparison, we choose the Dresselhaus constant $\gamma =0,40,80,120$
${\rm {eV}}$ ${\rm {{\AA}^{3}}}$.

We see the current fluctuation of the system caused by Dresselhaus
SOC. Fig. 7 presents results of the electric current $I$ and the
shot noise $S$ versus the external bias. It is demonstrated that the
peaks of the current and of the shot noise are lowered as the
Dresselhaus constant $\gamma$ increases, and a concave down in the
ascending side of the noise curve is obvious for non-zero $\gamma$.
The concaveness gives rise to nadirs far below 0.5 in the Fano
factor (see Fig. 8 (a)). To compare with experiment, $S$ vs $I$
curves normalized to unity are displayed in Fig. 8 (b). In the
positive differential conductance region, the shot noise follows the
value of uncorrelated electrons ($2eI$) for small tunnel currents
and is significantly suppressed for larger currents and eventually
increases. The suppression is above one-half for $\gamma=0$, near
one-half for $ \gamma  = 4.0 \times 10^{ - 29}$ ${\rm{eV}}$
${{\rm{m}}^{\rm{3}}} $, and below one-half for $\gamma$ larger than
$ 8.0 \times 10^{ - 29}$ ${\rm{eV}}$ ${{\rm{m}}^{\rm{3}}} $. In the
negative differential conductance (NDC) region, the Coulomb
interaction or charging effect in the well enhances the shot noise
and overweighs the effect of SOC\cite{Ref59}. Iannaccone et al. have
focused on the NDC region and obtained enhanced shot
noise\cite{Ref36}.

The results shown in Figs. 7-8 can be understood from the following.
The SOC interaction behaves like a pseudo magnetic field and induces
split of different spin components of the resonant level in the
barrier structure\cite{Ref55}, which contribute collectively to the
electric current and shot noise. Thus, the current is lowered at the
peak and simultaneously lifted in both sides around the peak in the
current-bias spectra. When the SOC is present, spin ``$\uparrow$"
electrons exclude spin ``$\downarrow$" electrons as well as spin
``$\uparrow$" ones. Therefore, the large noise suppressions are a
consequence of the repulsion between current pulses of different
spin states in addition to the consequence of the Pauli blockade and
Coulomb repulsion.

\section{Quantum pumping beyond linear response}

\subsection{Introduction to quantum pumping}

Generally speaking, the transport of matter from low potential to
high potential excited by absorbing energy from the environment can
be described as a pump process. The driving mechanics of classic
pumps is straightforward and well understood\cite{Ref61}. The
concept of a quantum pump is initiated several decades
ago\cite{Ref62} with its mechanism involving coherent tunneling and
quantum interference. Research on quantum pumping has attracted
heated interest since its experimental realization in an open
quantum dot\cite{Ref1, Ref2, Ref3, Ref4, Ref5, Ref6, Ref7, Ref8,
Ref9, Ref10, Ref11, Ref12, Ref13, Ref14, Ref15, Ref16, Ref17, Ref18,
Ref19, Ref63, Ref64, Ref65, Ref66, Ref67, Ref68, Ref69, Ref70,
Ref71, Ref72}.

The mechanisms of an adiabatic quantum pump can be demonstrated in a
mesoscopic system modulated by two oscillating barriers (see Fig.
9). To prominently picture the charge flow driven process within a
cyclic period, the two potential barriers are modulated with a phase
difference of $ \pi /2$ in the manner of $U_1 =U_0 + U_{1 \omega }
\sin t$ and $U_2 =U_0 + U_{2 \omega } \sin (t+ \pi /2)$. Our
discussion is within the framework of the single electron
approximation and coherent tunneling. The Pauli principle is taken
into account throughout the pumping process. The Fermi energy of the
two reservoirs and the inner single-particle state energy are
equalized to eliminate the external bias and secure energy-conserved
tunneling [The kinetic properties (charge current, heat current,
etc.) depend on the values of the scattering matrix within the
energy interval of the order of ${\rm{max}}(k_B T,\hbar \omega)$
near the Fermi energy. In the low-frequency ($\omega \rightarrow 0$)
and low-temperature ($T \rightarrow 0$) limit we assume the
scattering matrix to be energy independent]. As shown in Fig. 9, the
transmission strengths between one of the reservoirs and the inner
single-particle state are denoted by $t_1$-$t_4$. When $t \in [0,
\pi /2]$, $\sin t$ changes from 0 to 1 and $\sin (t+ \pi /2)$
changes from 1 to 0. Considering the time-averaged effect, the
chance of $U_1 >U_2$ and $U_1 <U_2$ is equal. Therefore, the
probability of $t_1$ and $t_3$ balance out. The tunneling quantified
by $t_2$ and $t_4$ do not occur since the inner particle state is
not occupied. When $t \in [\pi /2,\pi ]$, $\sin t$ changes from 1 to
0 and $\sin (t+ \pi /2)$ changes from 0 to -1. $U_1
>U_2$ invariably holds in this time regime. The probability of $t_3$
prevails and a net particle flow is driven from the right reservoir
to the middle state. When $t \in [\pi , 3 \pi /2]$, $\sin t$ changes
from 0 to -1 and $\sin (t+ \pi /2)$ changes from -1 to 0. The
probability of $t_2$ and $t_4$ balance out and the tunneling
quantified by $t_1$ and $t_3$ are excluded from the Pauli principle.
No net time-averaged tunneling occurs. When $t \in [3 \pi /2,2 \pi
]$, $\sin t$ changes from -1 to 0 and $\sin (t+ \pi /2)$ changes
from 0 to 1. $U_1$ maintains a lower height than $U_2$, which drives
the particle in the inner state to the left reservoir. Through one
whole pumping cycle, electrons are pumped from the right reservoir
to the left by absorbing energy from the two oscillating sources.
The tunneling is governed by quantum coherence. In each period, the
pumping process repeats and the particles are driven continuously in
the same direction as time accumulates. Direction-reversed pumped
current can be obtained with reversed phase difference of the two
oscillating gates. The direction of the pumped current is from the
phase-leading gate to the phase-lagged one without exception when we
assume that higher barriers admit smaller transmission probability.
It can find resemblance in its classical turnstile
counterpart\cite{Ref61} in which the fore-opened gate admits
transmission ahead of the later-opened one driving currents in
corresponding manner.

The current and noise properties in various quantum pump structures
and devices were investigated such as the magnetic-barrier-modulated
two dimensional electron gas\cite{Ref4}, mesoscopic one-dimensional
wire\cite{Ref6, Ref65}, quantum-dot structures\cite{Ref61, Ref5,
Ref11, Ref12, Ref71}, mesoscopic rings with Aharonov-Casher and
Aharonov-Bohm effect\cite{Ref7}, magnetic tunnel
junctions\cite{Ref10}, chains of tunnel-coupled metallic
islands\cite{Ref68}, the nanoscale helical wire\cite{Ref69},the
Tomonaga-Luttinger liquid\cite{Ref67}, and garphene-based
devices\cite{Ref63, Ref64}. Theory also predicts that charge can be
pumped by oscillating one parameter in particular quantum
configurations\cite{Ref66}. A recent experiment\cite{Ref70} based on
two parallel quantized charge pumps offers a way forward to the
potential application of quantum pumping in quantum information
processing, the generation of single photons in pairs and bunches,
neural networking, and the development of a quantum standard for
electrical current. Correspondingly, theoretical techniques have
been put forward for the treatment of the quantum pumps\cite{Ref2,
Ref3, Ref18, Ref65, Ref68, Ref72}. One of the most prominent is the
scattering approach proposed by Brouwer who presented
 a formula that relates the pumped current to the
parametric derivatives of the scattering matrix of the system.
Driven by adiabatic and weak modulation (the ac driving amplitude is
small compared to the static potential), the pumped current was
found to vary in a sinusoidal manner as a function of the phase
difference between the two oscillating potentials. It increases
linearly with the frequency in line with experimental finding. The
Floquet scattering theory is developed\cite{Ref72} for
quantum-mechanical pumping in mesoscopic conductors. It can be used
to investigate quantum pumping behavior at arbitrary pumping
amplitude and frequency.

As an example to demonstrate the Floquet scattering theory, we focus
on the experimentally observed deviation from the weak-pumping
theory with only the first-order parametric derivative of the
scattering matrix considered. By expanding the scattering matrix to
higher orders of the time and modulation amplitude, experimental
observation can be interpreted by multi-energy-quantum-related
processes.

\subsection{Theoretical formulation}

We use the scattering matrix approach to describe the response of a
mesoscopic phase-coherent sample to two slowly oscillating (with a
frequency $\omega$) external real parameters $X_{j}(t)$ (gate
potential, magnetic flux, etc.),
\begin{equation}
\begin{array}{*{20}c}
   {X_j \left( t \right) = X_{0,j}  + X_{\omega ,j} e^{i\left( {\omega t - \varphi _j } \right)}  + X_{\omega ,j} e^{ - i\left( {\omega t - \varphi _j } \right)} ,} & {j = 1,2.}  \\
\end{array}
\end{equation}
$X_{0,j}$ and $X_{\omega ,j}$ measure the static magnitude and ac
driving amplitude of the two parameters, respectively. The phase
difference between the two drivers is defined as $\phi  = \varphi _1
- \varphi _2 $. The mesoscopic conductor is connected to two
reservoirs at zero bias. The scattering matrix $\hat s$ being a
function of parameters $X_{j}(t)$ depends on time.

A time-dependent scattering matrix can be introduced as follows.
\begin{equation}
\hat b_\alpha  \left( t \right) = \int_{ - \infty }^\infty
{S_{\alpha \beta } \left( {t,t'} \right)\hat a_\beta  \left( {t'}
\right)dt'} ,\begin{array}{*{20}c}
   {} & {t \ge t'.}  \\
\end{array}
\end{equation}
Its Wigner transform reads
\begin{equation}
S_{\alpha \beta } \left( {E,t} \right) = \int_{ - \infty }^\infty
{e^{iE\left( {t - t'} \right)} S_{\alpha \beta } \left( {t,t'}
\right)dt'}.
\end{equation}

We assume the scattering time $t-t'$ is small. Up to corrections of
order $\hbar \omega / \gamma$ ($\gamma$ measures the escape rate),
the matrix $S_{\alpha \beta } \left( {E,t} \right)$ is equal to the
``instantaneous" scattering matrix $S_X (E)$, which is obtained by
``freezing" all parameters $X_j$ to their values at time $t$. Below,
we use the instant scattering matrix $\hat s\left( t \right)$ in
place of $S_{\alpha \beta } \left( {E,t} \right)$ to describe the
physics for simplicity. The kinetic properties (charge current, heat
current, etc.) depend on the values of the scattering matrix within
the energy interval of the order of ${\rm{max}}(k_B T,\hbar \omega)$
near the Fermi energy. In the low-frequency ($\omega \rightarrow 0$)
and low-temperature ($T \rightarrow 0$) limit we assume the
scattering matrix to be energy independent. To investigate the
deviation from the small amplitude $X_{\omega ,j} $ limit, we expand
the scattering matrix $\hat s (t)$ into Taylor series of $X_{j} (t)$
to second order at $X_{0,j}$ with the terms linear and quadratic of
$X_{\omega ,j} $ present in the expansion,
\begin{equation}
\hat s\left( t \right) \approx \hat s_{0}\left( {X_{0,j} } \right) +
\hat s_{ - \omega } e^{i\omega t}  + \hat s_{ + \omega } e^{ -
i\omega t}  + \hat s_2  + \hat s_{ - 2\omega } e^{2i\omega t}  +
\hat s_{ + 2\omega } e^{ - 2i\omega t} ,
\end{equation}
with
\begin{equation}
\left\{ \begin{array}{l}
 \hat s_{ \pm \omega }  = \sum\limits_{j = 1,2} {X_{\omega ,j} e^{ \pm i\varphi _j } {{\partial \hat s} \mathord{\left/
 {\vphantom {{\partial \hat s} {\partial X_j }}} \right.
 \kern-\nulldelimiterspace} {\partial X_j }}} , \\
 \hat s_2  = \sum\limits_{j = 1,2} {X_{\omega ,j}^2 {{\partial ^2 \hat s} \mathord{\left/
 {\vphantom {{\partial ^2 \hat s} {\partial X_j^2 }}} \right.
 \kern-\nulldelimiterspace} {\partial X_j^2 }}} , \\
 \hat s_{ \pm 2\omega }  = \frac{1}{2}\sum\limits_{j = 1,2} {X_{\omega ,j}^2 e^{ \pm 2i\varphi _j } {{\partial ^2 \hat s} \mathord{\left/
 {\vphantom {{\partial ^2 \hat s} {\partial X_j^2 }}} \right.
 \kern-\nulldelimiterspace} {\partial X_j^2 }}} . \\
 \end{array} \right.
\end{equation}
It can be seen from the equations that higher orders of the Fourier
spectra enter into the scattering matrix. As a result, both the
nearest and next nearest sidebands are taken into account, which
implies that a scattered electron can absorb or emit an energy
quantum of $\hbar \omega $ or $2 \hbar \omega $ before it leaves the
scattering region. In principle, third or higher orders in the
Taylor series can be obtained accordingly. However, the higher-order
parametric derivatives of the scatter matrix diminish dramatically
and approximate zero. Numerical calculation demonstrates that even
in relatively large amplitude modulation, their contribution is
negligible.

The pumped current depends on the values of the scattering matrix
within the energy interval of the order of $\max \left( {k_B T,2
\hbar \omega } \right)$ near the Fermi energy. In the
low-temperature limit ($T \to 0$), an energy interval of $2 \hbar
\omega $ is opened during the scattering process.

The mesoscopic scatterer is coupled to two reservoirs with the same
temperatures $T$ and electrochemical potentials $\mu$. Electrons
with the energy $E$ entering the scatterer are described by the
Fermi distribution function $f_{0} (E)$, which approximates a step
function at a low temperature. Due to the interaction with an
oscillating scatterer, an electron can absorb or emit energy quanta
that changes the distribution function. A single transverse channel
in one of the leads is considered. Applying the hypothesis of an
instant scattering, the scattering matrix connecting the incoming
and outgoing states can be written as
\begin{equation}
\hat b_\alpha  \left( t \right) = \sum\limits_\beta  {s_{\alpha
\beta } \left( t \right)\hat a_\beta  \left( t \right)}.
\end{equation}
Here $s_{\alpha \beta } $ is an element of the scattering matrix
$\hat s$; the time-dependent operator is $\hat a_\alpha  \left( t
\right) = \int {dE\hat a_\alpha  \left( E \right)e^{{{ - iEt}
\mathord{\left/
 {\vphantom {{ - iEt} \hbar }} \right.
 \kern-\nulldelimiterspace} \hbar }} } $,
and the energy-dependent operator ${\hat a_\alpha  \left( E
\right)}$ annihilates particles with total energy E incident from
the $\alpha$ lead into the scatter and obey the following
anticommutation relations
\begin{equation}
\left[ {\hat a_\alpha ^\dag  \left( E \right),\hat a_\beta  \left(
{E'} \right)} \right] = \delta _{\alpha \beta } \delta \left( {E -
E'} \right).
\end{equation}
Note that above expressions correspond to single- (transverse)
channel leads and spinless electrons. For the case of many-channel
leads each lead index ($\alpha $, $\beta$, etc.) includes a
transverse channel index and any repeating lead index implies
implicitly a summation over all the transverse channels in the lead.
Similarly an electron spin can be taken into account.

Using Eqs. (34) and (36) and after a Fourier transformation we
obtain
\begin{equation}
\begin{array}{l}
 \hat b_\alpha  \left( E \right) = \sum\limits_\beta  {\left[ {\hat s_{0,\alpha \beta } \hat a_\beta  \left( E \right) + \hat s_{2,\alpha \beta } \hat a_\beta  \left( E \right) + \hat s_{ - \omega ,\alpha \beta } \hat a_\beta  \left( {E + \hbar \omega } \right)} \right.}  \\
 \left. { + \hat s_{ + \omega ,\alpha \beta } \hat a_\beta  \left( {E - \hbar \omega } \right) + \hat s_{ - 2\omega ,\alpha \beta } \hat a_\beta  \left( {E + 2\hbar \omega } \right) + \hat s_{ + 2\omega ,\alpha \beta } \hat a_\beta  \left( {E - 2\hbar \omega } \right)} \right]. \\
 \end{array}
\end{equation}
The distribution function for electrons leaving the scatterer
through the lead $\alpha$ is $f_\alpha ^{\left( {out} \right)}
\left( E \right) = \left\langle {\hat b_\alpha ^\dag  \left( E
\right)\hat b_\alpha  \left( E \right)} \right\rangle $, where
$\left\langle  \cdots  \right\rangle $ means quantum-mechanical
averaging. Substituting Eq. (38) we find
\begin{equation}
\begin{array}{l}
 f_\alpha ^{\left( {out} \right)} \left( E \right) = \sum\limits_\beta  {\left[ {\left| {\hat s_{0,\alpha \beta }  + \hat s_{2,\alpha \beta } } \right|^2 f_0 \left( E \right) + \left| {\hat s_{ - \omega ,\alpha \beta } } \right|^2 f_0 \left( {E + \hbar \omega } \right)} \right.}  \\
 \left. {\left| {\hat s_{ + \omega ,\alpha \beta } } \right|^2 f_0 \left( {E - \hbar \omega } \right) + \left| {\hat s_{ - 2\omega ,\alpha \beta } } \right|^2 f_0 \left( {E + 2\hbar \omega } \right) + \left| {\hat s_{ + 2\omega ,\alpha \beta } } \right|^2 f_0 \left( {E - 2\hbar \omega } \right)} \right]. \\
 \end{array}
\end{equation}
The distribution function for outgoing carriers is a nonequilibrium
distribution function generated by the nonstationary scatterer. The
Fourier amplitudes of the scattering matrix ${\left| {\hat s_{ -
\omega ,\alpha \beta } } \right|^2 }$ (${\left| {\hat s_{ + \omega
,\alpha \beta } } \right|^2 }$) is the probability for an electron
entering the scatterer through the lead $\beta$ and leaving the
scatterer through the lead $\alpha$ to emit (to absorb) an energy
quantum $\hbar \omega $ and ${\left| {\hat s_{ - 2\omega ,\alpha
\beta } } \right|^2 }$ (${\left| {\hat s_{ + 2\omega ,\alpha \beta }
} \right|^2 }$) is that of the energy quantum $2 \hbar \omega $
process. ${\left| {\hat s_{0,\alpha \beta }  + \hat s_{2,\alpha
\beta } } \right|^2 }$ is the probability for the same scattering
without the change of an energy with the second-order term $\hat
s_{2,\alpha \beta }$ much smaller than the zero-order term $\hat
s_{0,\alpha \beta }$ in weak-modulation limit ($X_{\omega ,j}  \ll
X_{0,j} $) and can be omitted therein.

Using the distribution functions $f_{0} (E)$ for incoming electrons
and $f_{\alpha} ^{out} (E)$ for outgoing electrons, the pumped
current measured at lead $\alpha$ reads
\begin{equation}
I_p  = \frac{e}{{2\pi \hbar }}\int_0^\infty  {\left\langle {\hat
b_\alpha ^\dag  \left( E \right)\hat b_\alpha  \left( E \right)}
\right\rangle  - \left\langle {\hat a_\alpha ^\dag  \left( E
\right)\hat a_\alpha  \left( E \right)} \right\rangle dE}.
\end{equation}
Substituting Eqs. (34) and (30) we get
\begin{equation}
\begin{array}{c}
 I_p  = \frac{{e\omega }}{{2\pi }}\sum\limits_{\beta ,j_1 ,j_2 } {X_{\omega ,j_1 } X_{\omega ,j_2 } \frac{{\partial s_{\alpha \beta } }}{{\partial X_{j_1 } }}\frac{{\partial s_{\alpha \beta }^* }}{{\partial X_{j_2 } }}2i\sin \left( {\varphi _{j_1 }  - \varphi _{j_2 } } \right)}  \\
  + \frac{{e\omega }}{{2\pi }}\sum\limits_{\beta ,j_1 ,j_2 } {X_{\omega ,j_1 }^2 X_{\omega ,j_2 }^2 \frac{{\partial ^2 s_{\alpha \beta } }}{{\partial X_{j_1 }^2 }}\frac{{\partial ^2 s_{\alpha \beta }^* }}{{\partial X_{j_2 }^2 }}i\sin \left[ {2\left( {\varphi _{j_1 }  - \varphi _{j_2 } } \right)} \right]} . \\
 \end{array}
\end{equation}
Quantum pumping properties beyond the theory based on first-order
parametric derivative of the scattering matrix are demonstrated in
Eq. (41). By taking higher orders of the Fourier spectrum of the
scattering matrix into consideration, double $\hbar \omega $ energy
quantum (or a $2 \hbar \omega $ energy quantum) emission
(absorption) processes coact with single $\hbar \omega $ quantum
processes. In the weak-modulation limit, the second term in the
right-hand side of Eq. (41) is small, which implies that double
$\hbar \omega $ quantum processes are weak and therefore not
observable. As the ac driving amplitude is enlarged, this term
increases markedly and contribution from double $\hbar \omega $
quantum processes takes effect. As a result, the dependence of the
pumped current on the phase difference between two driving
oscillations deviates from sinusoidal and changes from $\sin \phi $
to $\sin 2\phi $, which is observed in experiment\cite{Ref1}.
Moreover, the relation between the pumped current and the ac driving
amplitude $X_{\omega ,j} $ is reshaped. It is also seen that the
linear dependence of the pumped current on the oscillation frequency
holds for multi-quanta-related processes.

\subsection{Numerical results and interpretations}

Here, numerical results of the pumped current in a
two-oscillating-potential-barrier modulated nanowire are presented
and comparison with experiment is given. We consider a nanowire
modulated by two gate potential barriers with equal width $L=20$
{\AA} separated by a $2L=40$ {\AA} width well (see Fig. 10). The
electrochemical potential of the two reservoirs $\mu$ is set to be
$60$ meV according to the resonant level within the double-barrier
structure. The two oscillating parameters in Eq. (31) correspond to
the two ac driven potential gates $X_{1,2} \left( t \right) \to
U_{1,2} \left( t \right)$ with all the other notations correspond
accordingly. We set the static magnitude of the two gate potentials
$U_{0,1}  = U_{0,2} = U_0  = 100$ meV and the ac driving amplitude
of the modulations equal $U_{\omega ,1}  = U_{\omega ,2} = U_\omega
$.

In Fig. 11, the dependence of the pumped current on the phase
difference between the two ac oscillations is presented. In
weak-modulation regime (namely $U_\omega  $ is small), sinusoidal
behavior dominates. Here, three relatively large $U_\omega  $ is
selected to reveal the deviation from the sinusoidal dependence.
(The magnitude of the pumped current mounts up in power-law relation
as a function of $U_\omega  $ as shown in Fig. 12. The sinusoidal
curve for small $U_\omega  $ would be flat and invisible in the same
coordinate range.) It can be seen from the figure that the $I_p$-$
\phi $ relation varies from sinusoidal ($\sin \phi $) to
double-sinusoidal ($\sin 2 \phi $) as the ac oscillation amplitude
is increased. The interpretation follows from Eq. (41). The single
$\hbar \omega $ quantum emission (absortion) processes feature a
sinusoidal behavior while the $2 \hbar \omega $ quantum emission
(absortion) processes feature a double-sinusoidal behavior when the
Fourier index is doubled. As $U_\omega  $ is increased, double
$\hbar \omega $ quantum processes gradually parallel and outweigh
the single $\hbar \omega $ quantum ones. It is also demonstrated
that when the single $\hbar \omega $ quantum processes have the
effect of $\sin \phi $ dependence, the double $\hbar \omega $
quantum processes induce a $-\sin 2 \phi $ contribution with a sign
flip, which can be understood from the sign change of the derivative
of the scattering matrix. The effect of three- and higher $\hbar
\omega $ quantum processes is small even for large $U_\omega $
comparable to $U_0$. The experimental observations\cite{Ref1} as a
deviation from the weak-modulation limit are revealed by our theory.

Experiment\cite{Ref1} also discovered that for weak pumping the
dependence of the pumped current on the pumping strength obeys a
power of 2 relation, as expected from the simple loop-area
argument\cite{Ref2}; for strong pumping, power of 1 and 1/2 relation
is observed. We presented in Fig. 12 the numerical results based on
our theory of the $I_p$-$U_{\omega}$ relation at a fixed $\phi $. To
demonstrate its power-law dependence, natural logarithm of
 the variables is applied. From Eq. (41), it can be seen that for
large ac driving amplitude $U_{\omega}$, contribution of double
$\hbar \omega $ quantum processes (formulated in the second term on
the right hand side of the equation) causes the $I_p$-$U_{\omega}$
relation to deviate from its weak-modulation limit, the latter of
which is $I_p \propto U_\omega ^2 $. For different phase difference
between the two ac drivers, the deviation is different. At $\phi
=\pi $ the pumped current is invariably zero regardless of the order
of approximation determined by time-reversal symmetry. At $\phi =\pi
/2$, $\sin 2\phi $ is exact zero, and no difference is incurred by
introducing higher order effect. If we shift the value of $\phi$ to
$0.49 \pi$, the abating effect of the double $\hbar \omega $ quantum
processes has the order of $U_\omega ^4 $ with the small
second-order parametric derivative of the scattering matrix
smoothing that effect a bit. Consequently, a power of $2 \to 1 \to
1/2$ relation is obtained and visualized by the curve fit, which is
analogous to experimental findings. For different values of $\phi$,
sharper abating and augmental effect occurs with analogous
mechanisms. It is possible that the experiment\cite{Ref1} was done
at the phase difference close to $\pi /2$ while trying to approach
maximal pumped current in the adiabatic and weak-pumping limit.

\section{Summary and future directions}

The scattering matrix method is initiated by Landauer and
B\"{u}ttiker to investigate the conductance of multi-terminal and
multi-channel mesoscopic conductors. The spin degree of freedom can
be included in the formalism by enlargement of the dimension of the
scattering matrix. The current-current correlation and spin-spin
correlation, such as the shot noise, can be calculated from the
cross products of the scattering matrix. Along this direction,
higher-order correlation function can also be considered. Dynamic
transport processes including the quantum pumping behavior can be
dealt with by the time-dependent scattering approach.

The development of the scattering theory enables its potential
applications in currently open issues. The interaction can be
included to the scattering matrix by the renormalization factors.
The shot noise properties in various conductors with active spin
degree of freedom can be considered. The time-dependent scattering
theory provides a way to deal with dynamic quantum issues. Some
particular problems include non-harmonically driven quantum pumping,
spin pumping in racetrack memory applications, and multiferroic
transport dynamics, etc..

\section{Acknowledgements}

This project was supported by the Nature Science Foundation of SCUT
(No. x2lxE5090410) and the Graduate Course Construction Project of
SCUT (No. yjzk2009001). The author would like to express sincere
appreciation to Professor Wenji Deng, Dr. Brian M. Walsh, Professor
Jamal Berakdar, Professor Michael Moskalets, and Professor Liliana
Arrachea for valuable enlightenment to the topic from discussions
with them.

\clearpage

\begin{figure}[t]
 \includegraphics{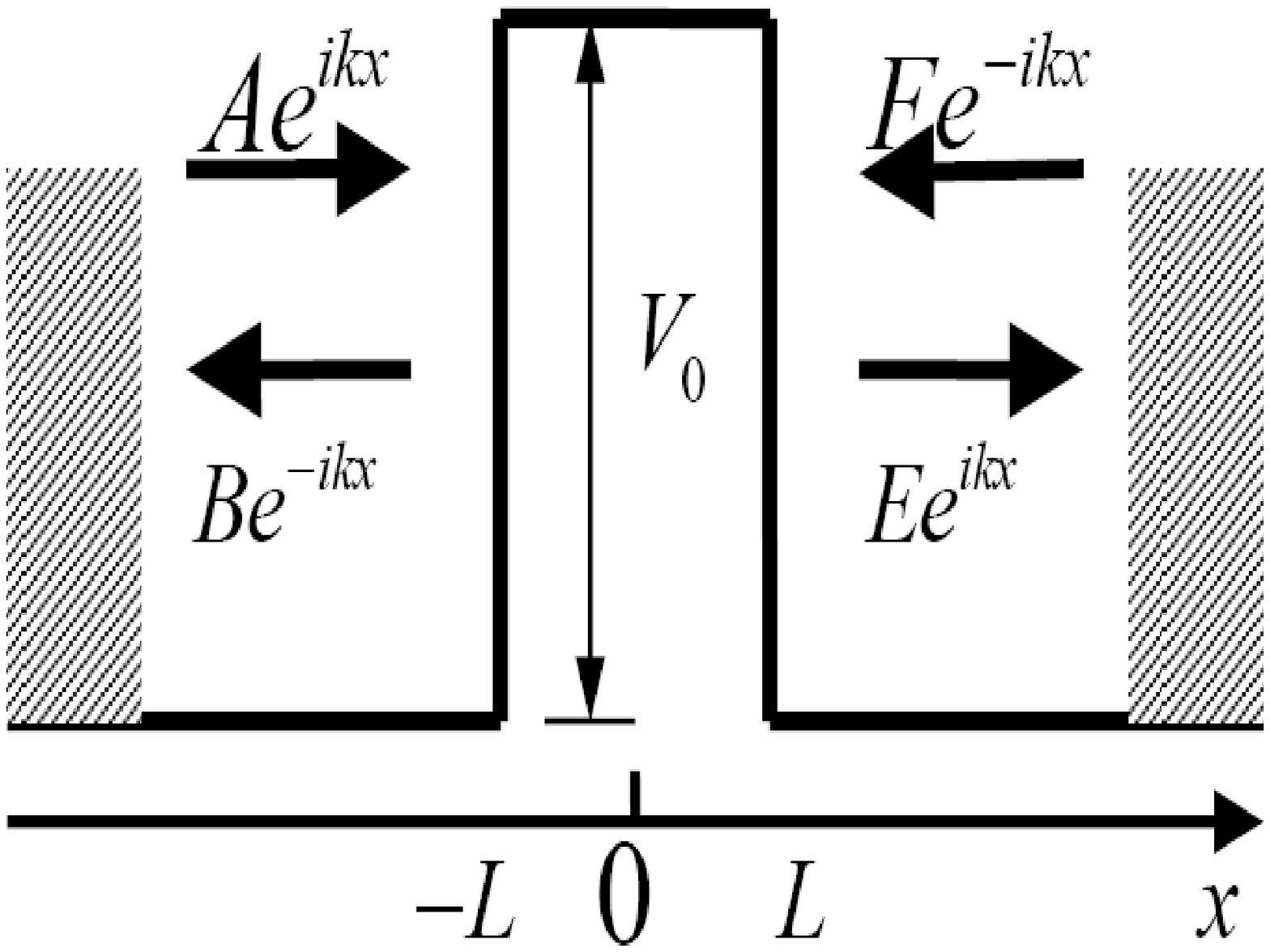}
\caption{Schematic demonstration of a single-barrier tunneling
problem. The quantum states between two reservoirs at zero bias is
indicated. }
\end{figure}

\begin{figure}[h]
 \includegraphics{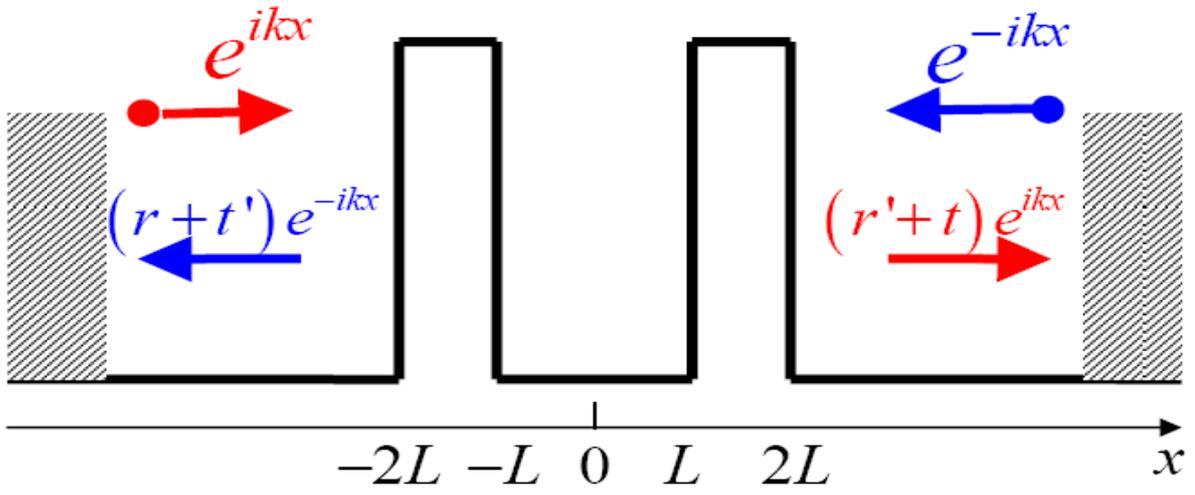}
\caption{Schematic demonstration of a quantum wire modulated by two
potential barriers. The quantum state between two reservoirs at zero
bias is indicated.}
\end{figure}

\begin{figure}[h]
 \includegraphics{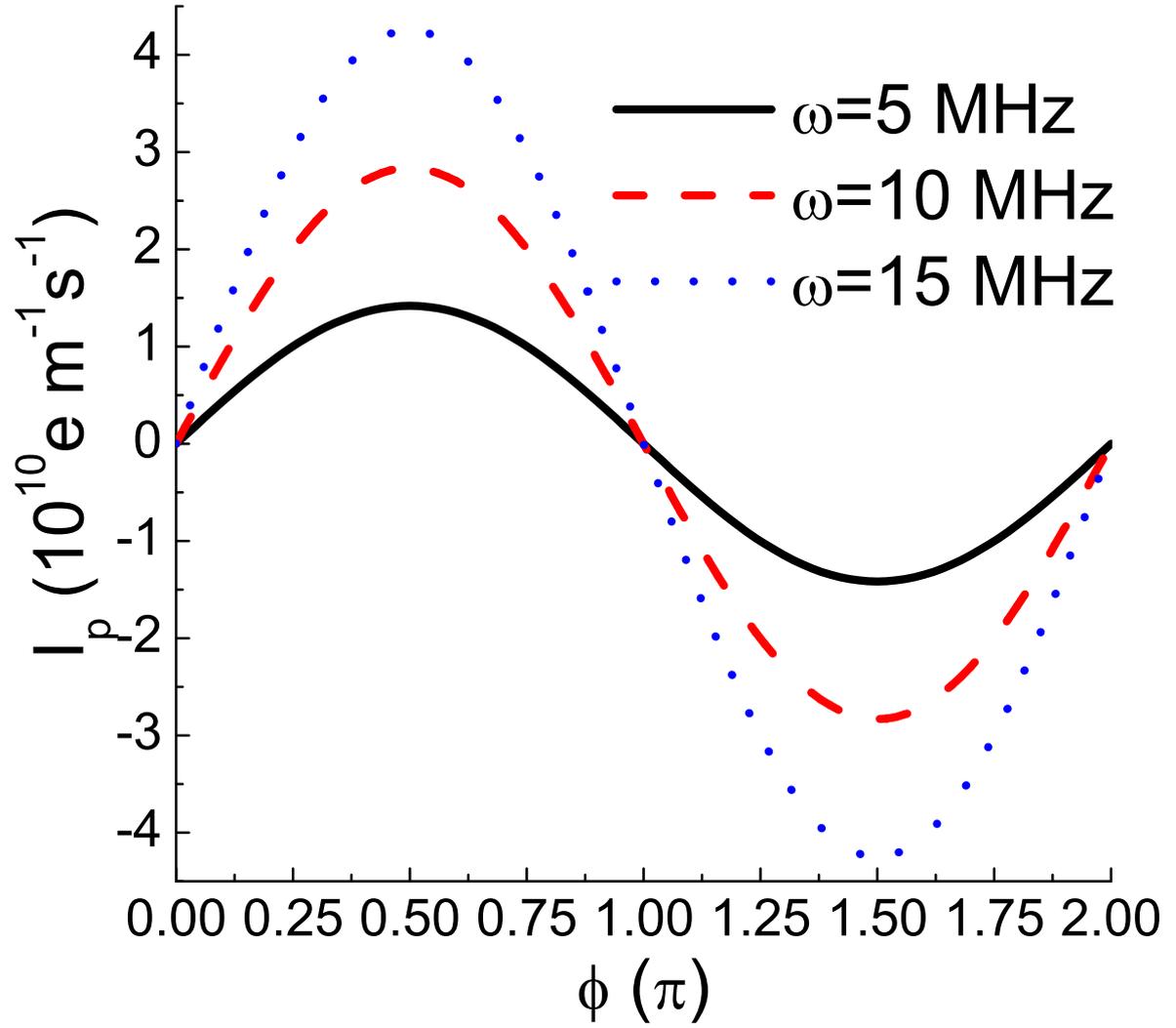}
\caption{Time-integrated current density as a function of the phase
difference
 between the two modulations for different modulation frequencies. The
 Fermi level of the two reservoirs $E_F=60$ meV counting from the
 conduction band edge of the electron gas structure.}
\end{figure}

\begin{figure}[h]
 \includegraphics{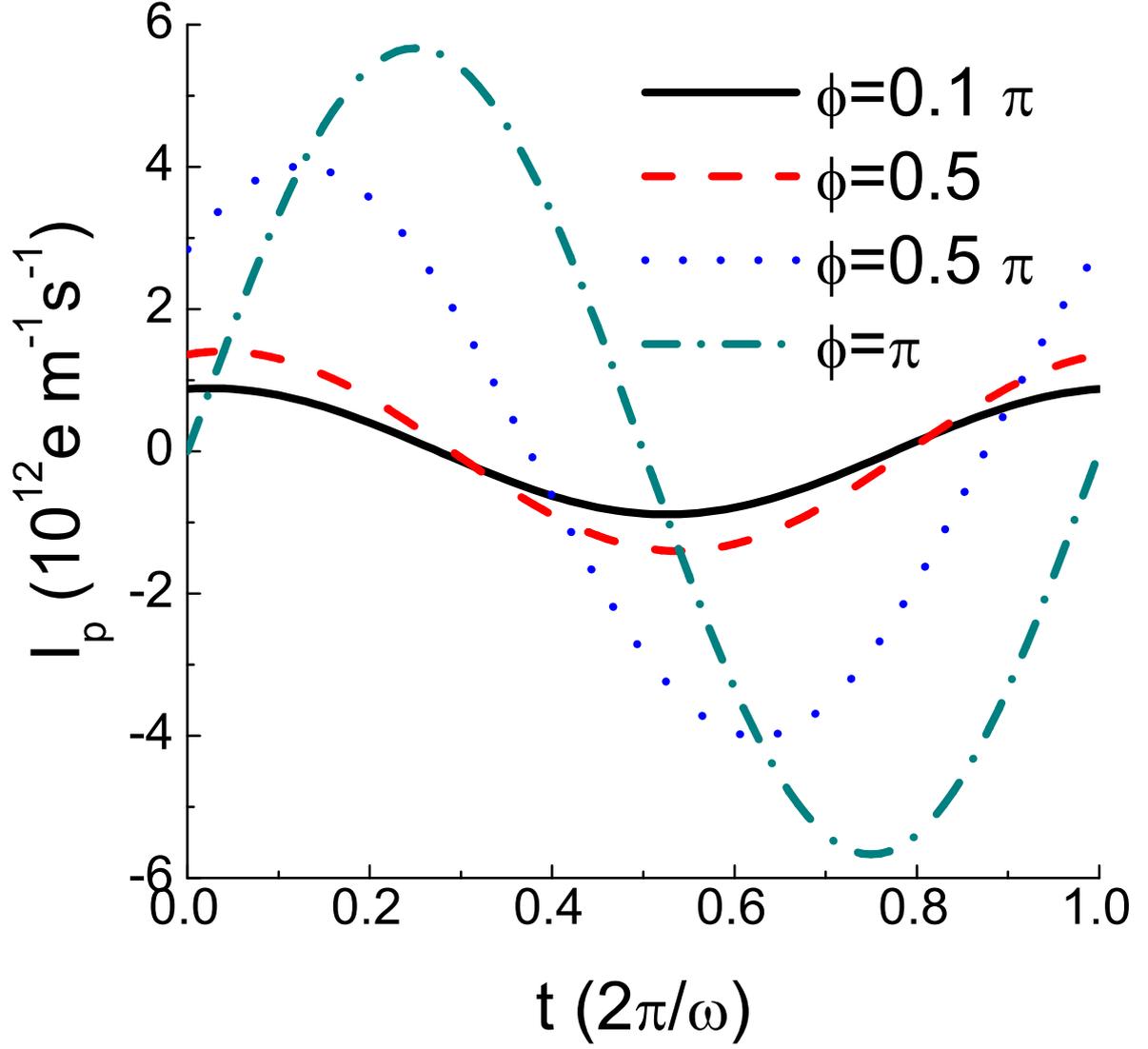}
\caption{Time variation of the current density within
 an oscillating cycle. Different phase difference between the two
 modulations is considered. The modulation frequency is
 set to be $\omega  = 10$ MHz and the Fermi level $E_F$ to be 60 meV.}
\end{figure}

\begin{figure}[h]
 \includegraphics[width=11cm]{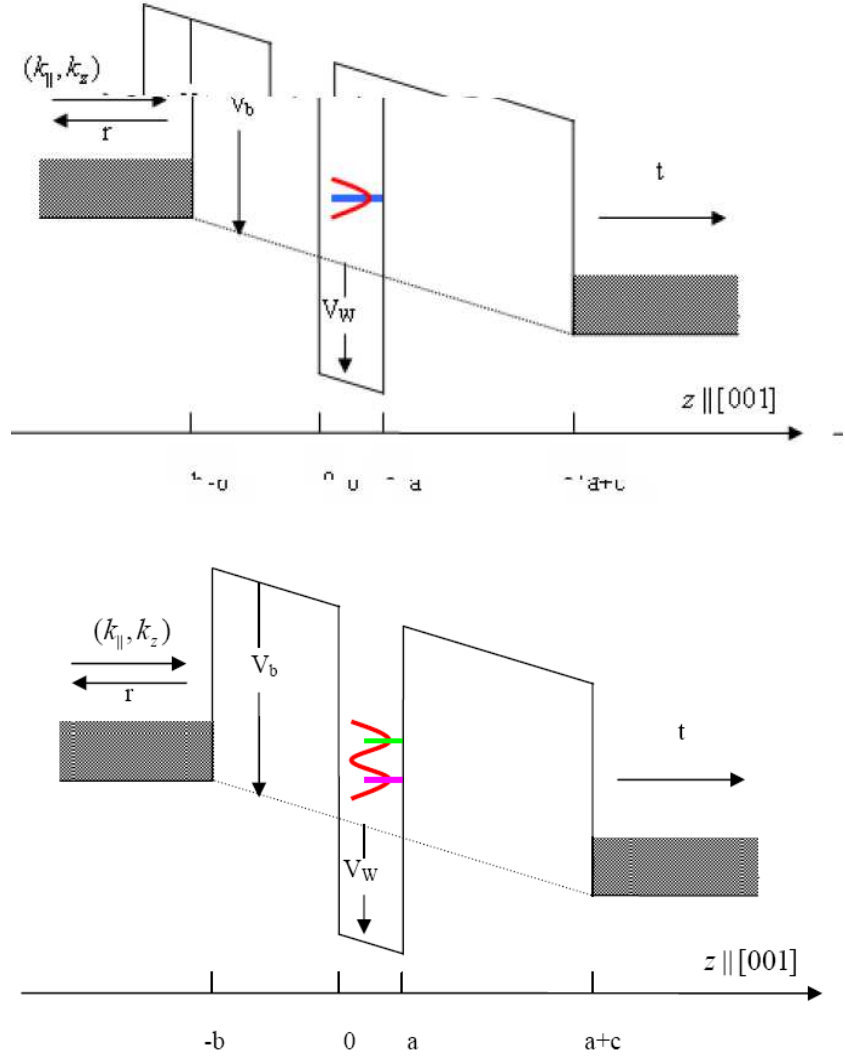}
\caption{Schematics of the double-barrier resonant diode. The
resonant level is sketched between the two barriers. The upper panel
demonstrates the resonant level in conventional diode without the
spin-orbit coupling (SOC). In the lower panel, the SOC behaves like
a pseudomagnetic field and induces a split of different spin
components of the resonant level in the barrier structure, which
contribute collectively to the electric current and shot noise.}
\end{figure}

\begin{figure}[h]
 \includegraphics{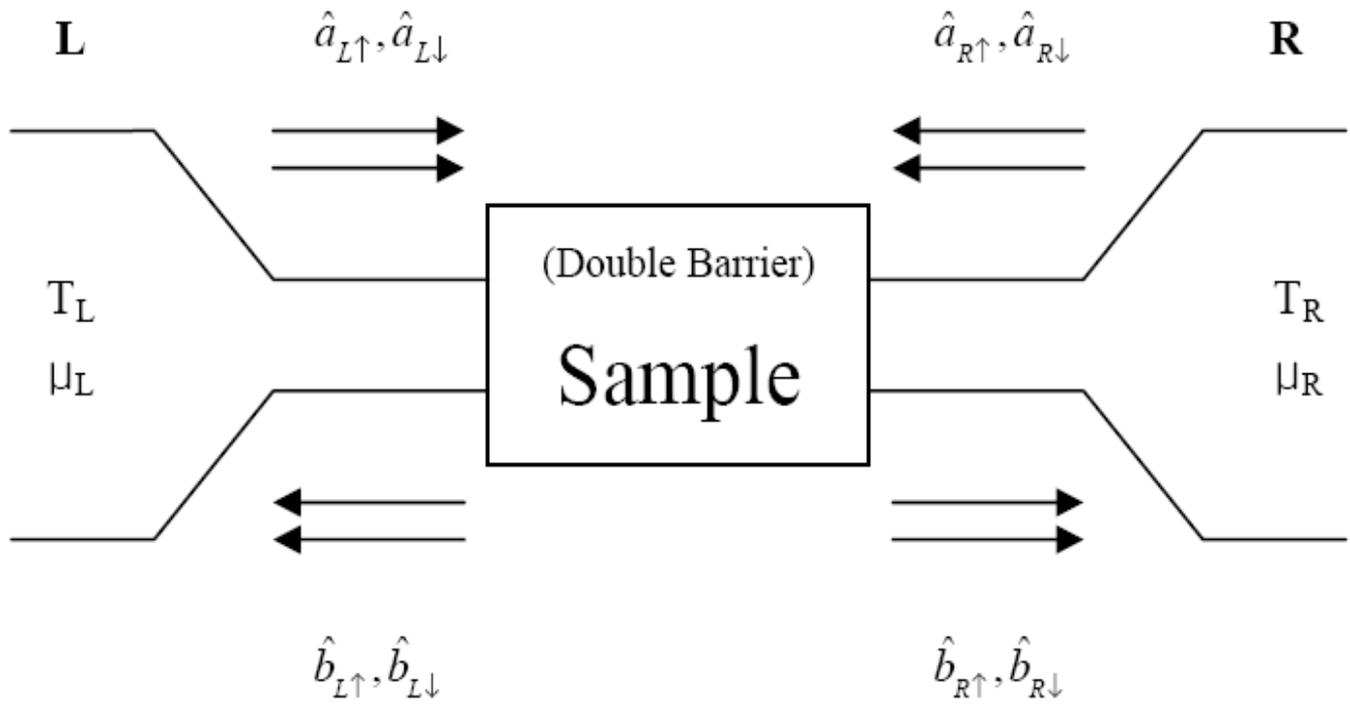}
\caption{Schematics of the scattering approach. Spin components of
the incoming and outgoing states are indicated.}
\end{figure}

\begin{figure}[h]
 \includegraphics{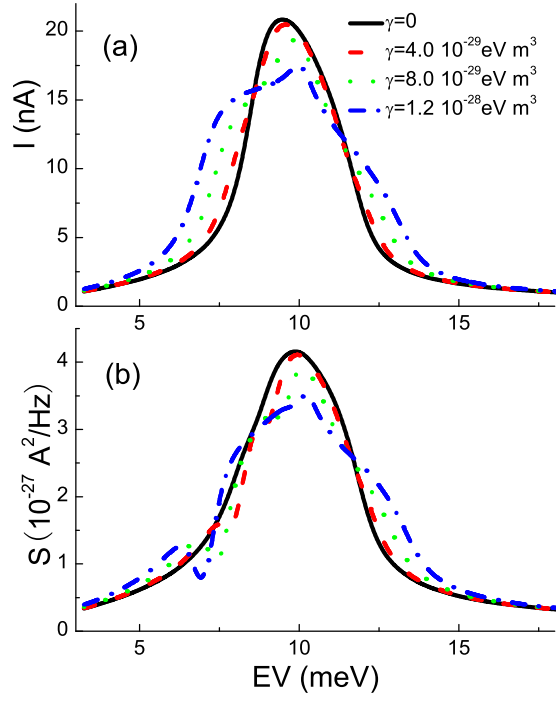}
\caption{Current $I$ (a) and shot noise $S$ (b) as functions of the
applied bias $EV$ of electrons traversing a symmetric DBRD structure
with different Dresselhaus constants $\gamma$.}
\end{figure}

\begin{figure}[h]
 \includegraphics{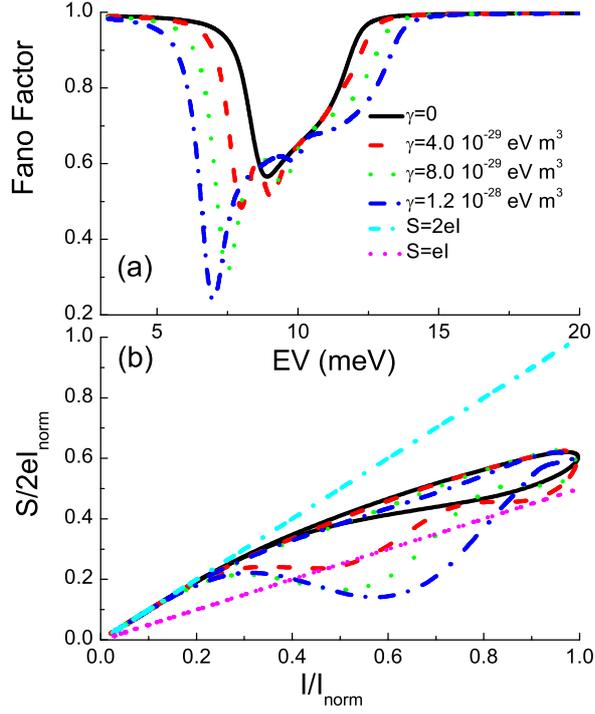}
\caption{Fano factor as a function of the applied bias $EV$ (a) and
$S/2eI_{norm}$ vs $I/I_{norm}$ (b) of electrons traversing a
symmetric DBRD structure with different Dresselhaus constants
$\gamma$. Two straight lines in (b) show the full shot noise value
($2eI$) and half of its value for comparison.}
\end{figure}

\begin{figure}[h]
 \includegraphics[width=11cm]{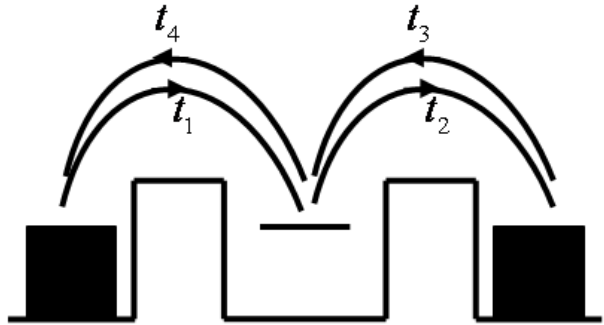}
\caption{The tunneling
 scenario of an adiabatic quantum pump. The two shadowed blocks
 represent the left and right electron reservoirs respectively. The two barriers oscillate adiabatically in
 time. The middle bar indicates the
 single-particle state between the two barriers. The
 Fermi levels of the two reservoirs are the same and are leveled to the
 single-particle state within the conductor.
  $t_1$-$t_4$ indicate the transmission
 amplitudes between one of the two reservoirs and the middle
 single-particle state.}
\end{figure}

\begin{figure}[h]
 \includegraphics{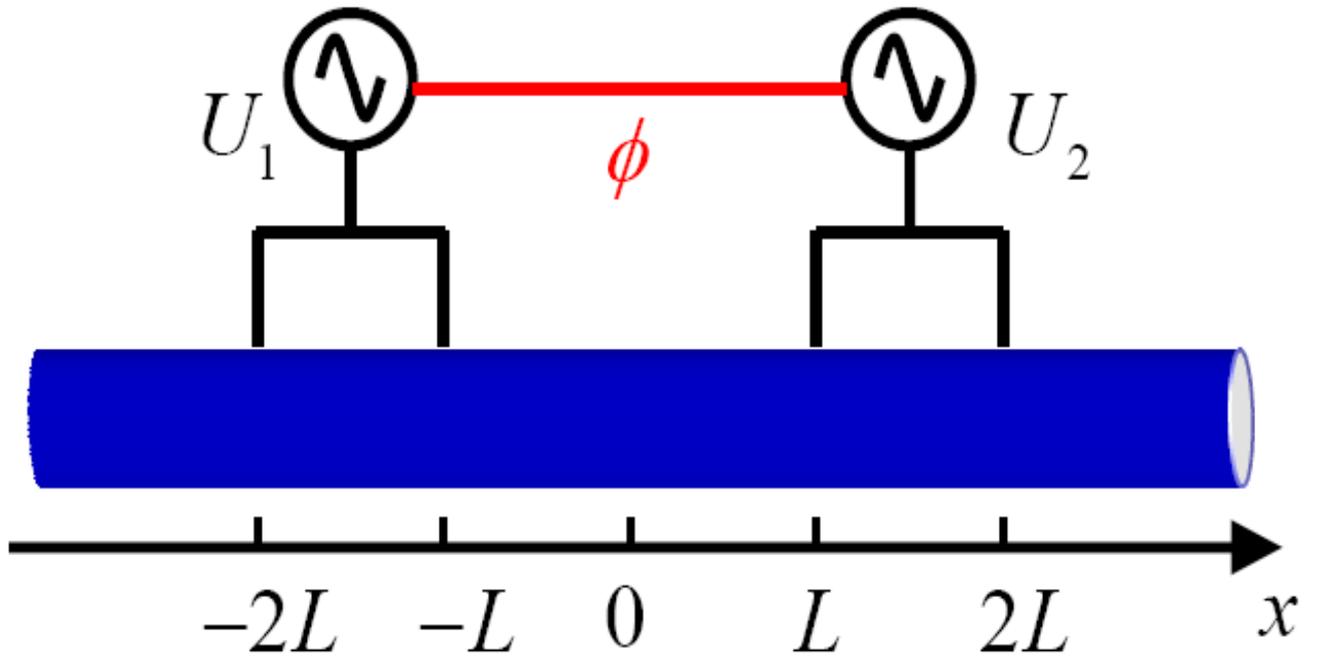}
\caption{Schematics of the quantum pump: a nanowire modulated by two
ac driven potential barriers.}
\end{figure}

\begin{figure}[h]
 \includegraphics{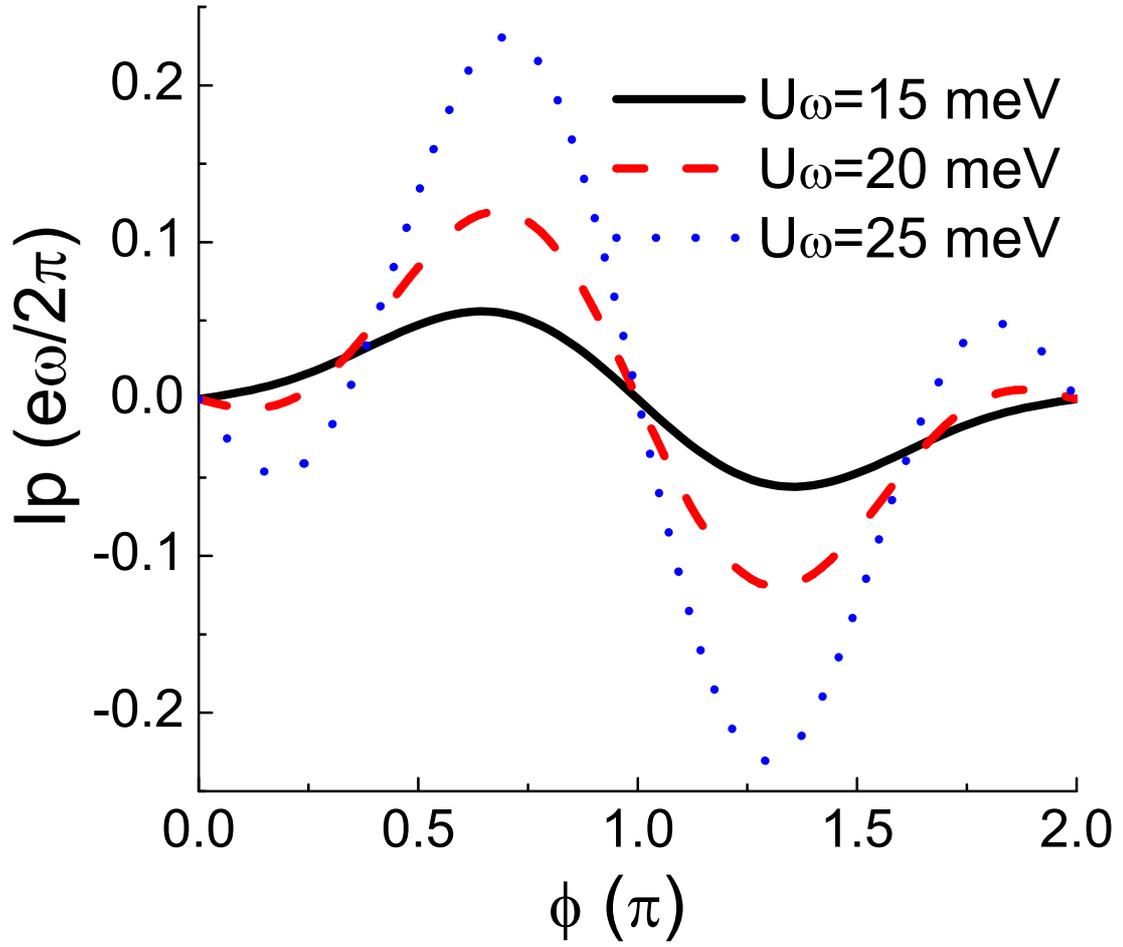}
\caption{Pumped current as a function of the phase difference
 between the two modulations for different ac driving amplitudes.}
\end{figure}

\begin{figure}[h]
 \includegraphics{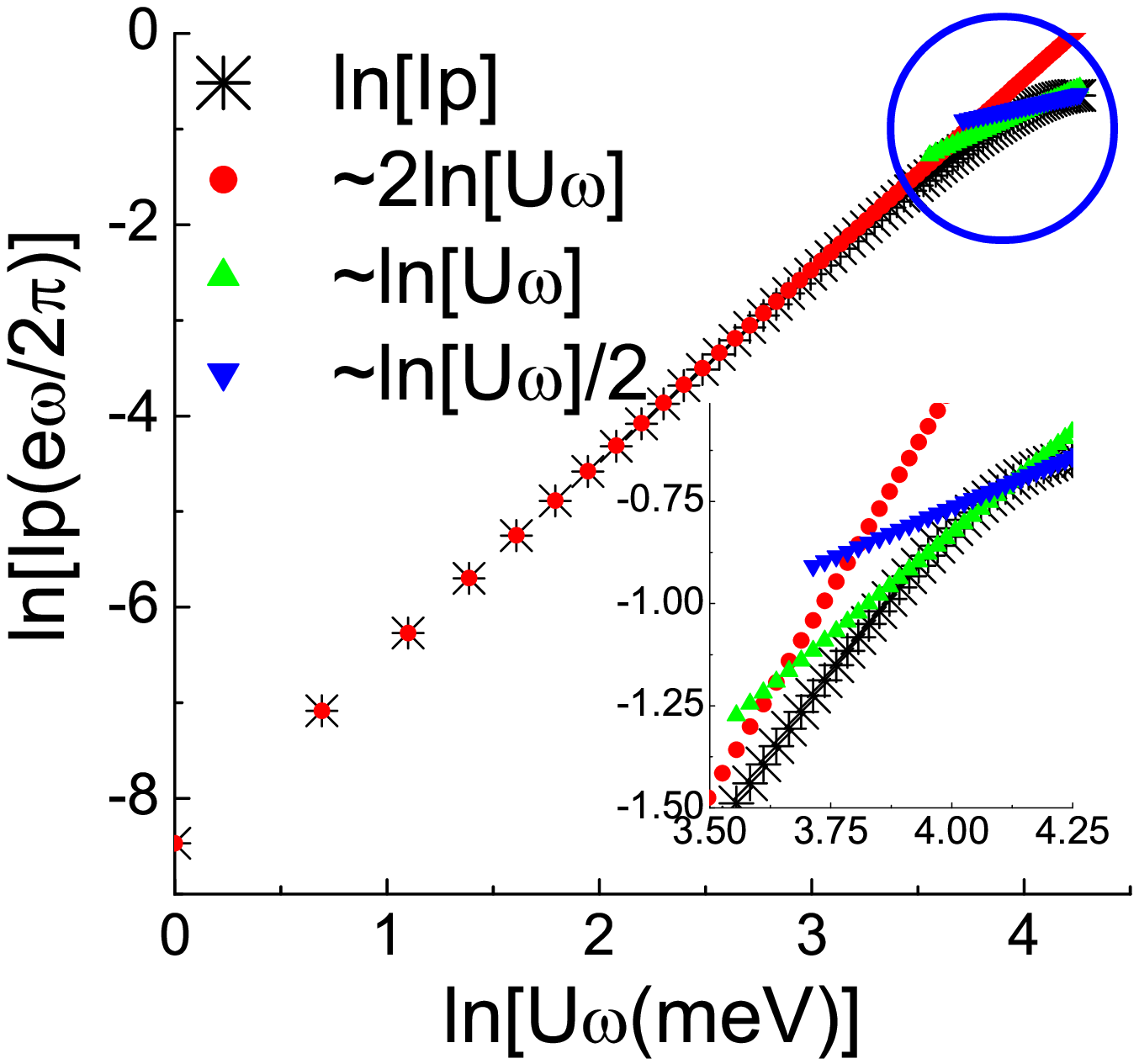}
\caption{Pumped current as a function of the ac driving amplitude
$U_\omega
 $ along with fits to $I_p  \propto U_\omega ^2 $ (red solid
 circle) below 35 meV,
$I_p  \propto U_\omega  $ (green upward triangle) below 41 meV, and
$I_p \propto U_\omega ^{1/2} $ above 41 meV (blue downward
triangle). To demonstrate its power-law dependence, natural
logarithm of
 the variables is applied.
 The phase difference between the two ac driver $\phi  = 0.49\pi $.
 Inset is the zoom-in of the circled region.}
\end{figure}


\begin{references}
\bibitem{Ref59} In the positive differential conductance (PDC) region,
results for the case considering charging effect or Coulomb
interaction are close to those neglecting Coulomb effect in both
coherent and sequential approaches (see Refs. \onlinecite{Ref36} and
\onlinecite{Ref60}), i.e., the effect of Coulomb interaction on shot
noise is weak in the region. Thus it is reasonable to neglect it in
the PDC region.
\end{references}
\end{document}